# The nucleosynthetic fingerprint of the outermost protoplanetary disk and early Solar System dynamics

Elishevah van Kooten[1*], Xuchao Zhao[2], Ian Franchi[2], Po-Yen Tung[3], Simon Fairclough[3], John Walmsley[3], Isaac Onyett[1], Martin Schiller[1], Martin Bizzarro[1,4]

[1] Center for Star and Planet Formation, Globe Institute, University of Copenhagen; Øster Voldgade 5-7, 1350 Copenhagen, Denmark
[2] School of Physical Sciences, Open University, Milton Keynes, MK7 6AA, UK
[3] Department of Materials Science and Metallurgy; 27 Charles Babbage Road, Cambridge, CB3 0FS
[4] Institut de Physique du Globe de Paris, Université de Paris, 1 Rue Jussieu, 75005 Paris, France.

*Corresponding author: elishevah.vankooten@sund.ku.dk

**Abstract:** Knowledge of the nucleosynthetic isotope composition of the outermost protoplanetary disk is critical to understand the formation and early dynamical evolution of the Solar System. We report the discovery of outer disk material preserved in a pristine meteorite based on its chemical composition, organic-rich petrology, and $^{15}N$-rich, deuterium-rich, and $^{16}O$-poor isotope signatures. We infer that this outer disk material originated in the comet-forming region. The nucleosynthetic Fe, Mg, Si and Cr compositions of this material reveal that, contrary to current belief, the isotope signature of the comet-forming region is ubiquitous amongst outer Solar System bodies, possibly reflecting an important planetary building block in the outer Solar System. This nucleosynthetic component represents fresh material added to the outer disk by late accretion streamers connected to the ambient molecular cloud. Our results show that most Solar System carbonaceous asteroids accreted material from the comet-forming region, a signature lacking in the terrestrial planet region.

**Short title:** The nucleosynthetic fingerprint of comets

**One-Sentence Summary:** The cometary nucleosynthetic component is ubiquitous amongst outer Solar System asteroids and planets.

## INTRODUCTION

The most primitive material in the Solar System is thought to have initially resided in the cold, outer regions of the protoplanetary disk during the early formative stages of the Solar System. This includes cometary bodies, which are composed of ice, organics, and rocky material derived from the Kuiper belt located >30 astronomical units (a.u.) from the Sun and the more distant Oort cloud. Although comets have been traditionally viewed as compositionally distinct from asteroids, it is now well-established that a compositional continuum exists between these bodies such that comets may represent an endmember composition of the Solar System's planetary building blocks. Comets are not thought to be represented in meteorite collections, but some astromaterials such as chondritic porous interplanetary dust particles (CP-IDPs) and Ultracarbonaceous Antarctic micrometeorites likely originate from the comet-forming region (*1–4*). Detailed investigations of these IDPs and micrometeorites have provided insights into the petrology, chemical make-up, and isotope composition of light elements such as H, N and O characteristic of cometary material.

Knowledge on the formation and early evolution of the Solar System comes from primitive meteorites known as carbonaceous and non-carbonaceous chondrites. Unlike comets, these meteorites are sourced from asteroids accreted either in the terrestrial planet-forming region (non-carbonaceous) or close to the gas giant planets (carbonaceous) (*5*). Together with Earth and Mars, these asteroidal fragments define large-scale nucleosynthetic isotope variability for heavy elements, interpreted to reflect heterogeneous distribution of isotopically anomalous stardust in the protoplanetary disk (*6*). A significant cosmochemical observation is the existence of a gradient in the nucleosynthetic composition of planetary materials with orbital distance (*7*). Material accreted relatively close to the protosun like Earth, Mars and the parent asteroids of non-carbonaceous meteorites has a different nucleosynthetic make-up relative to the water-rich carbonaceous asteroids (*6–8*). The origin of this compositional contrast is debated, and several models have been forward, including thermal processing and selective destruction of stardust (*7*) as well as primordial heterogeneity of the Solar System's parental molecular cloud core (*9*). Distinguishing between these models is central to test models of Solar System formation and understand the dynamical evolution of the protoplanetary disk.

Identifying the nucleosynthetic fingerprint of material formed in the outermost protoplanetary disk and, by extension, the comet-forming region, will provide insight into the origin of the Solar System's nucleosynthetic variability. This is because current models predict contrasting compositions for the outermost protoplanetary disk, including cometary material. In one view, the nucleosynthetic composition of this disk region should be akin to that of primordial molecular cloud material (*10*) whereas the competing view predicts a composition like that of the first Solar System solids (*9*), calcium-aluminium-rich refractory inclusions (CAIs). Available astromaterials such as cometary IDPs and micrometeorites could allow us to determine the nucleosynthetic fingerprint of cometary material, but their small size prevents extracting meaningful nucleosynthetic data. A source of outer disk material may be preserved in metal-rich carbonaceous chondrites, which include the CR (Renazzo-like), CB (Bencubbin-like) and CH (high metal) chondrites. Their Mg and Cr isotope composition suggest a genetic link to comets (*10*). Moreover, in addition to potential outer disk xenolithic candidates from non-carbonaceous chondrites (*11–16*), material of outer disk origin has been previously identified in a CR chondrite (*17*). Thus, we conducted a systematic search for outer disk material in the primitive CR chondrite North West Africa (NWA) 14250, which contains abundant mm- to cm-sized dark clasts (i.e., pebble-shaped, chondritic material unrelated to the host body) (*18*). Detailed chemistry, petrology as well as H, N, C, and O isotope systematics of several dark clasts establish that these objects are derived from the outermost protoplanetary disk, the region where most cometary bodies accreted. This provides the opportunity to determine the nucleosynthetic fingerprint of the comet-forming region and, as such, unravel the accretion history of the solar protoplanetary disk.

**RESULTS AND DISCUSSION**

**Preservation of interstellar matter in the dark clasts**

The $Cr_2O_3$ content of Fe-rich chondrule olivine functions as a thermometer that probes a meteorite's thermal history (*19*, *20*). Figure S1 shows that NWA 14250 is amongst the most pristine CR chondrites based on this thermometer. Thus, the dark clasts in NWA 14250 may represent the most pristine outer Solar System material available, including thermally labile organics present in these clasts (see SM, Table S1 and Fig. S1).

The dark clasts are dominated by fine-grained submicron-sized matter (resembling chondritic matrix and CI chondrites) and the larger clasts contain at maximum 10 vol.% of chondrules (Fig. 1; Fig. S2-6). Small chondrules fragments were also observed in comet Wild2 (*4*) and resemble the FeO-rich microchondrules present in the dark clasts (see SM, Fig. S2-S4). The clast matrix has a solar composition, determined by comparing the chemical composition of the clasts for elements recording different condensation temperatures with CI chondrites and samples from the Cb-asteroid Ryugu (Fig. 1). The clast petrography, including the high abundance of fine-grained matrix, approximates that of the Tarda and Tagish Lake chondrites, which are suggested to derive from trans-Neptunian D-type asteroids (*21*, *22*). Thus, the CI-like chemical composition is not restricted to CI chondrites but reflects a ubiquitous outer Solar System dust reservoir (*23*).

Nanoscale characterization by transmission electron microscopy (TEM, see Methods) highlights the low abundance of hydrated relative to amorphous silicates (Fig. 2, Fig S7-11), establishing that the clast matrix experienced a very low degree of aqueous alteration. The matrix is dominated by sulfide- and organic-rich matter (~40 vol.%, Fig. 2) interspersed with Si-rich areas consisting of amorphous to finely crystalline silicates. The sulfide and organic complexes (SOC) are not completely devoid of Si, suggesting the presence of amorphous silicate in the organic-rich area. Nanoscale sulfides and Mg-rich carbonates are restricted to the SOC areas and interpreted as primary condensates (see SM for detailed explanation). These SOC resemble glass with embedded metal and sulfide phases (i.e., GEMS) in IDPs and GEMS-like materials in primitive chondrites (*24, 25*). The absence of metal and their relative sizes place the origin of these SOC most closely to GEMS-like matter in primitive chondrites (see SM). However, their organic-rich nature, including their structural relationship to cometary IDPs (Fig. 2G and H) (*26*) and anomalous isotope signatures (see next section), exclude a high temperature condensation origin in the solar nebula (*24, 27, 28*) and promote the origin of SOC in cold environments like the outer Solar System (*29*) or the ISM (*30*).

**An outer disk origin for the dark clasts**

Matrix areas within the dark clasts were analyzed for their N, H, C and O isotopes by nano secondary ion mass spectrometry (nanoSIMS) (Figs. 2 and 3 and Data S2). Multiple isotope maps (17 for N and O isotopes and 8 for H and C isotopes) from three clasts collectively define identical bulk $\delta^{15}$N, $\delta$D and $\delta^{13}$C compositions of 378±32 ‰, 274±74 ‰ and −30±17 ‰ (2SD), respectively. The fourth clast (C3DC4) has lower $\delta^{15}$N and $\delta$D values, like hydrated dark clasts previously observed in the Isheyevo meteorite (*31*) (Fig. S12C-D). We focus our discussion on the least altered matrix areas within the three clasts with highly anomalous bulk $\delta^{15}$N, $\delta$D and $\delta^{13}$C values. We show in Figure 3 the distribution of $^{15}$N, D and $^{13}$C in the bulk maps, the carbon-rich regions and the most $^{15}$N- and D-rich regions. We observe a general decrease in $^{15}$N with increasing carbon content (using $^{12}C^{14}N$ as a proxy) and show that the most $^{15}$N-rich regions essentially coat the carbon-rich material. In contrast, the $\delta$D values increase with increasing carbon content (using $^{12}C$ as a proxy) and the most D-rich regions fall within the carbon-rich areas, whereas the $\delta^{13}$C values appear to decrease with carbon content.

Figure 4A shows the $\delta^{15}$N and $\delta$D values of the bulk composition of the dark clasts plotted together with that of bulk carbonaceous chondrites and IDPs. The bulk composition of the dark clasts is similar to cometary IDPs and the C2 chondrite Bells, which are amongst the most isotopically anomalous Solar System materials. We show in Figure 4B the average bulk clast composition together with that of the C-rich and the most D-rich areas (from Fig. 3). These data points form a mixing line between two endmembers. One endmember is relatively D-rich, $^{13}$C-depleted and $^{15}$N-depleted organic matter that falls on the line predicted for organic matter derived from cold ion-molecule reactions in the ISM (*30*).

Moreover, this correlated D-rich and $^{13}C$-depleted composition is also consistent with H and C isotope trends from chemical reaction networks in dense prestellar cores (*32*). The second endmember is D-poor and $^{15}N$-rich and could reflect the composition of the original ice accreted to the dark clast parent body (Fig. 4B) that was subsequently molten and present as a hydrating fluid. The $\delta D$ value of this water-ice component is $-206\pm120$ ‰ (see Fig. S13), in agreement with previous measurements of carbonaceous chondrite initial water (*31, 33*) ($\delta D_{initial} = -350$ ‰ to $+100$ ‰). The high $\delta^{15}N$ and low $\delta D$ values of the altering fluid (i.e., the original ice composition) can be explained if the original ice predominantly consisted of $H_2O$, with <1% $NH_3$ with a $\delta^{15}N$ >1000 ‰ (Table S2).

It is unlikely that the large $^{15}N$-enrichments (>1000 ‰) observed in objects such as Isheyevo lithic clasts (*31, 34, 35*), the Zag clast (*11, 12*) and the dark clasts studied here are produced by low temperature ion-molecule reactions in dense clouds (*36, 37*). Instead, N isotope fractionation through photodissociation at the surface of protoplanetary disks, [self-shielding isotope effect (*38*)], can produce highly anomalous $^{15}N$ enrichments (*39*). Assuming a primary ISM-like N and H isotope composition for the ices, re-equilibration in the disk with a D-poor $H_2$-gas can decrease the $\delta D$ composition of ices to the observed values. Because vertical mixing of micron-sized particles is efficient in the outer disk (*40*), photodissociation of $N_2$ gas at the disk surface can increase the $\delta^{15}N$ values of the icy grains. Thus, the inferred high $\delta^{15}N$ and low $\delta D$ values of the ices accreted to the dark clasts are best explained by ice processing in the outer disk.

We interpret the N, H and C isotope composition of dark clast phases as ISM-derived micron-sized SOC (i.e., D-rich IOM endmember) coated by icy grains and processed in the outer disk. This unique mixing of primitive outer Solar System components has previously only been observed in anhydrous lithic clasts from Isheyevo (*31*) and provides a plausible mechanism to explain the H and N isotope composition of IDPs (Fig. 4A). The bulk $\delta^{15}N$ and $\delta D$ signatures of the dark clasts are best explained by a mixture of 40% organics and 60% $H_2O$ + $NH_3$ ice (Table S2 and Fig. S14). The combined H and N isotope ratios and petrology of the dark clasts suggest an ice : organic : silicate ratio similar to comets (*41*) (43 : 29 : 29, Table S3 and Fig. S15).

The bulk oxygen isotope composition of the dark clasts is very $^{16}O$-poor and falls along a slope 1 correlation line (*42*), with average $\delta^{17}O$ and $\delta^{18}O$ values (except C3DC4) of 26.2±4.3 ‰ and 24.7±5.5 ‰, respectively (Fig. 5) and a $\Delta^{17}O$ value of 14.3±8.5 (2SD). The slope 1 line reflects mixing between the $^{16}O$-rich solar composition with a $^{16}O$-poor ice endmember produced by CO self-shielding isotope effects in the outer disk (*38*) or in the protosolar molecular cloud (*43*). The oxygen isotope composition of the dark clasts is more $^{16}O$-poor than any bulk Solar System material, including the Balmoral bulk IDP and the Zag clast. The Balmoral bulk IDP is suggested to be of cometary origin (*44*) whereas the Zag clast is believed to be derived from P or D asteroids (*11, 12*). The most $^{16}O$-poor signatures are observed in hotspots from IDPs (*3, 45*) that plot near the inferred composition of primordial water ice derived from cosmic symplectites (i.e., intergrowths of magnetite and FeNi sulfides) from Acfer 094 (*46, 47*). That the clasts represent the most [16]O-poor bulk planetary material to date suggest incorporation of abundant primordial ice as expected for comets. We conclude that clasts A3_DC1, A3_DC2 and C2C3 are derived from the outermost protoplanetary disk and, as such, can be used as proxies to determine the nucleosynthetic composition of the comet-forming region.

**The nucleosynthetic fingerprint of the comet-forming region**

We measured the nucleosynthetic Fe, Mg, Si and Cr isotope composition of four dark clasts (Table S4 and S5, Fig. 6) of which two have been analysed for their detailed chemical and N, H, C and O isotope composition and are confirmed outer disk dark clasts (ODDs). Although the chemical composition of these clasts is CI-like, their nucleosynthetic compositions do not match CI chondrites. Focusing on the two confirmed ODDs, their average $\mu^{54}$Fe value of 33±3 ppm and $\mu^{54}$Cr of 125±9 ppm are distinct from CIs but identical to CR chondrites (*48, 49*). Their $\mu^{30}$Si of 4.7±1.8 ppm is distinct from any carbonaceous chondrite. The $\mu^{26}$Mg* value of these clasts defines a deficit of −20.3±0.4 ppm relative to CI chondrites (*50*), an isotope signal only observed in primitive components of metal-rich carbonaceous chondrites (*10*). Because both the clasts and CI chondrites have near-solar Al/Mg ratios, their distinct Mg isotope composition can only reflect precursor heterogeneity in either their initial $^{26}$Al/$^{27}$Al ratio or their Mg isotopes. We note that Mg and Cr isotope data has been published for C1-like (i.e., having experienced intensive aqueous alteration) carbonaceous dark clasts from polymict ureilites (*15, 16*). However, their isotope composition (Mg, Cr, O) suggests an affinity to CI chondrites and, thus, is distinct from the ODDs reported here. A key observation emerging from our data is that the nucleosynthetic composition of the dark clasts, and by extension, cometary materials, is close to that of metal-rich carbonaceous chondrites, suggesting a genetic link between comets and these asteroidal parent bodies. Moreover, our results establish that material with a solar chemical composition (i.e., CI-like dark clasts and CI chondrites) show a range of nucleosynthetic isotope signatures.

CI chondrites and samples from the Cb-asteroid Ryugu have identical nucleosynthetic Fe and Ti isotope compositions distinct from other carbonaceous chondrites (*51*). This has been interpreted as a common accretion region for CIs and Ryugu hypothesized to be the same as comets. This is not consistent with the Fe, Mg and Si nucleosynthetic isotope signatures from the ODDs (Fig. 6A). Moreover, we note that the H, N and O isotope composition of CI chondrites (*52–55*) is markedly different from that of IDPs and comets, which is incompatible with a common accretion region. In contrast, a genetic link between metal-rich carbonaceous chondrites and comets is supported by the IDP-like [15]N- and D-rich signatures of CR and CH/CB chondrite matrix (*31*, *34*, *35*, *56*) and O isotopes of CR chondrules that match those sampled by the Stardust mission (*57*).

Two contrasting models have been put forward to explain the Solar System's wide nucleosynthetic isotope heterogeneity. In one model the nucleosynthetic isotope contrast between non-carbonaceous and carbonaceous chondrites – also referred to as the dichotomy (*5*) – is ascribed to a compositional change of envelope material accreting to the protoplanetary disk in concert with outward transport of early accreted material by viscous expansion (*9*). The composition of the early accreted material is hypothesized to match that of CAIs (Fig. 6B). Following disk expansion resulting in an isotopically CAI-like disk, the inner disk (i.e., <3 a.u.) is replenished by late addition of non-carbonaceous matter. Thus, this model predicts a CAI-like nucleosynthetic isotope signature for the outermost disk, including the comet-forming region (*58*). The nucleosynthetic fingerprint of the ODDs is not CAI-like, ruling out an early compositional change in the envelope material accreted to the inner disk as a cause of the Solar System's nucleosynthetic variability.

In the competing model, thermal processing of CI-like dust results in the unmixing of different nucleosynthetic carriers, namely a thermally labile supernova component from one that is galactically inherited and thermally robust (*7*, *50*) (Fig. 6B). The thermally processed inner disk represented by the non-carbonaceous composition is depleted in the supernova component, whereas CAIs reflect the complementary composition enriched in this component. In this model, the metal-rich carbonaceous chondrites represent a distinct nucleosynthetic signature restricted to the outermost protoplanetary disk (*10*). In $\mu^{26}$Mg\*-$\mu^{54}$Cr space, metal-rich carbonaceous chondrites and their components define nucleosynthetic variability accounted for by two endmembers, namely thermally processed dust and primordial molecular cloud material typified by a CI-like Cr isotope composition and a $\mu^{26}$Mg\* value (–15.9±1.4 ppm) representing the initial Solar System composition (*50*, *59*). The primordial molecular cloud material composition is interpreted to reflect the nucleosynthetic makeup of the molecular cloud before the last addition of stellar-derived $^{26}$Al. A corollary of this model is that primitive bodies accreted beyond Neptune should be dominated by the primordial molecular cloud endmember composition. This model is consistent with the average $\mu^{26}$Mg\* and $\mu^{54}$Cr values of the ODDs (Fig. 6B), which establish that these objects are dominated by primordial molecular cloud matter.

**Dynamical evolution of the protoplanetary disk**

Figure 6C shows the $\mu^{54}$Fe-$\mu^{30}$Si systematics of ODDs, CI, and other carbonaceous chondrites as well as non-carbonaceous bodies. Two distinct correlations are observed, namely one between CI and inner disk non-carbonaceous bodies and one between CI and outer disk carbonaceous bodies, including the ODDs. The inner disk correlation is ascribed to the progressive admixing of CI-like dust to a thermally processed inner Solar System reservoir (*49, 60*). In contrast, the outer disk correlation requires admixing of a distinct nucleosynthetic component represented by the ODDs to a CI-like reservoir to account for the variability observed across carbonaceous parent bodies. This implies that apart from CI, all carbonaceous chondrite groups incorporated variable proportions of material derived from the comet-forming region, suggesting that their accretion regions were not physically isolated. This observation emphasizes the existence of a compositional continuum between cometary bodies and outer disk asteroids. Because carbonaceous chondrites accreted late relative to inner Solar System bodies (*61*), the cometary nucleosynthetic component must represent a new addition of material to an initially CI-like disk. Astronomical observations (*62–64*) and numerical simulations (*65, 66*) indicate that late infall of molecular cloud material from accretion streamers is common in protoplanetary disks. Streamers are filamentary arms connecting the disk to its larger scale protostellar environment. Streamers can feed outer disk regions (>150 au) with fresh material from a different reservoir relative to the initial disk composition, which can provide a significant amount of mass after the class 0 phase of stellar evolution. The cometary nucleosynthetic component identified here may reflect that of the material accreted late to the outer disk, which we infer represents $^{26}$Al-free primordial molecular cloud material.

The absence of the cometary nucleosynthetic signature in the inner protoplanetary disk suggests that this component was accreted at large orbital distances. Indeed, the drift timescales of 100 µm-sized particles initially located at 100 AU to reach the inner disk is approximately 5 Myr (*67*), consistent with the lack of the cometary nucleosynthetic signature in NC bodies. The observation that carbonaceous parent bodies have accreted variable amounts of the cometary nucleosynthetic component supports an outer disk origin for this signature.

Accepting that the cometary nucleosynthetic component represents $^{26}$Al-free ambient molecular cloud material, our data favor the supernova-trigger hypothesis for the formation of the Solar System (*68*). In detail, the shockwave from a supernova event triggers the gravitational collapse and injection of gas and dust into the collapsing cloud core parental to the Solar System, thereby accounting for the former presence of $^{26}$Al and other extinct short-lived radionuclides (*69*). In this model, CI chondrites represent the bulk composition of the cloud core polluted by supernova-derived $^{26}$Al (e.g., early infall). In contrast, the cometary nucleosynthetic component represents the composition of the ambient molecular cloud unpolluted by a recent supernova event (e.g., late infall). As such, the Solar System trichotomy observed in $\mu^{30}$Si-$\mu^{54}$Fe space is the expression of three nucleosynthetic components, namely the bulk Solar System composition (CI chondrites), the primordial molecular cloud composition (material from the comet-forming region) and, the thermally processed inner disk (non-carbonaceous bodies) (Fig. 6C). Finally, since significant mass can be accreted via late infall, the bulk of the material located in the outer Solar System must be dominated by the cometary nucleosynthetic component. This can be tested by high-precision coupled $^{54}$Fe-$^{30}$Si measurements of cometary micrometeorites, newly defined and/or ungrouped carbonaceous chondrites and, lastly, samples returned from the Ryugu and Bennu primitive asteroids.

## MATERIALS AND METHODS

<u>Scanning electron microscopy (SEM) and laser ablation inductively coupled plasma mass spectrometry (LA-ICPMS)</u>
Several slabs of the CR chondrite NWA 14250 were mounted in 1-inch epoxy coins and polished down to submicron resolution. The dark clasts were identified and imaged in back scattered electron (BSE) mode by Zeiss EVO 15 scanning electron microscope at StarPlan (University of Copenhagen). EDS analyses were carried out on regions of interest to obtain quantitative major element concentrations using a MAC Universal Mineral standard block for reference. Laser ablation ICPMS analyses were carried out at the Department of Geology at Lund University. Carbon coatings were removed with ethanol from the thick sections before analysis. A 193 nm Cetac Analyte G2 excimer laser installed with a two volume HelEx2 sample cell was connected to a Bruker Aurora Elite ICP-MS. The laser was heated for at least 15 minutes before operation to ensure a stable laser output energy. The plasma of the ICPMS was lit minimum an hour before run time to stabilize the background signal. After cleaning, the samples were inserted into the ablation cell, which was flushed several times with the helium carrier gas to reduce the gas blank level. Data were acquired from single spot analysis of 60 μm, using a nominal laser fluence of ~2 J/cm$^2$ and a pulse rate of 10 Hz. The total acquisition time for a single analysis was approximately two minutes, including 30 s gas blank measurement followed by laser ablation for 63 s and washout for 20 s. All samples were measured using a standard-sample bracketing technique, where 8 samples were alternated with repeats of NIST610 and NIST612 glass standards and BHVO-2 and BIR-1 geological reference standards. The following isotopes were analyzed during all sessions: $^{23}$Na, $^{24}$Mg, $^{27}$Al, $^{29}$Si, $^{31}$P, $^{33}$S, $^{35}$Cl, $^{43}$Ca, $^{49}$Ti, $^{51}$V, $^{53}$Cr, $^{55}$Mn, $^{57}$Fe, $^{59}$Co, $^{60}$Ni, $^{65}$Cu, $^{66}$Zn, $^{71}$Ga, $^{72}$Ge, $^{77}$Se, $^{85}$Rb, $^{89}$Y, $^{90}$Zr, $^{93}$Nb, $^{95}$Mo, $^{114}$Cd, $^{118}$Sn, $^{125}$Te, $^{139}$La, $^{140}$Ce, $^{141}$Pr, $^{146}$Nd, $^{147}$Sm, $^{153}$Eu, $^{157}$Gd, $^{159}$Tb, $^{163}$Dy, $^{165}$Ho, $^{166}$Er, $^{169}$Tm, $^{172}$Yb, $^{175}$Lu, $^{178}$Hf, $^{182}$W, $^{205}$Tl, $^{208}$Pb, $^{238}$U. Si, also analyzed by SEM, was used as the internal standard element. Data reduction was performed off-line through the Iolite v.2.5 software and the trace element data reduction scheme (*70*). The limit of detection (LOD) was calculated using the 3-sigma deviation of the blanks. Four smaller dark clasts without chondrules and four larger clasts with chondrules were analyzed for their matrix composition using multiple 60 μm spots (n = 3 to 5) to calculate the average composition. We obtained three different averages for each clast, based on their correction against BHVO-2, BIR-1 and NIST612 standards. These averages were normalized against the CI chondrite composition and averaged to produce the final composition for a dark clast with a 2sd propagated uncertainty (see Data S1).

<u>Nanometer secondary ion mass spectrometry (nanoSIMS)</u>
The sections including SEM imaged dark clasts were transferred to the nanoSIMS laboratory at the Open University and were gold coated (20 nm thickness). Samples were pre-sputtered with a ~25 pA primary Cs$^+$ beam for ~15 min until sputter equilibrium was achieved. For analyses, a beam current of 1.0 pA was used, providing a spatial resolution of ~150 nm. The primary beam was rastered over 10 × 10 μm (256 × 256 pixels) with a dwell time of 2.5 ms/pixel, and we accumulated 20 repeated measurements on each area. The secondary ions of $^{16}$O$^-$, $^{17}$O$^-$, $^{18}$O$^-$, $^{12}$C$^{14}$N$^-$, $^{12}$C$^{15}$N$^-$ and $^{28}$Si$^-$ were measured in multi-collection mode together with the San Carlos olivine and an inhouse organic standard HOBt (*44*). For each of the four clasts (A3DC1, A3DC2, C2DC5 and C3DC4), 4-7 areas were mapped with this setup. Subsequently, 2-3 of these areas per clast were selected for additional analyses using a Cs$^+$ primary beam and including the secondary ions $^1$H$^-$, $^2$H$^-$, $^{12}$C$^-$, $^{13}$C$^-$ and $^{18}$O$^-$. The results were processed using L'image software (L. Nittler), including correction for detector deadtime (44 ns) and sample drift (Data S2). H, C and N isotope results are reported as δ$^{13}$C$_{PDB}$, δ$^{15}$N$_{AIR}$, δD$_{SMOW}$ and δ$^x$O$_{SMOW}$. A detailed description of the overall setup is given by Starkey and Franchi (*44*). We recognize that the use of San Carlos Olivine (SCO) standard for the correction of instrumental mass fractionation (IMF) in fine-grained matrix can induce matrix effects that can shift the data along a mass-dependent fractionation line (i.e., parallel to the TFL line in Fig. 5). We have calculated the ΔIMF between the SCO and a CI chondrite composition to assess this offset using the correction algorithm from Dubinina et al. (*71*), which we found to be -1.6 ‰ on $\delta^{18}O$. This effect is similar to mass bias offsets found between the SCO and serpentines with a range of compositions (<3 ‰), although mass bias in a SHRIMP instrument may not behave exactly the same as SIMS instruments (72). This small offset combined with the fact that our bulk oxygen isotope data plot on the slope 1 line, suggests that the matrix effects do not significantly affect our main conclusions. Finally, the IMF-inferred uncertainties are small relative to the uncertainties of our measurements and the observed offset in $\delta^{17}O$ and $\delta^{18}O$ from the TFL in the dark clasts.

Transmission electron microscopy (TEM)

Dark clast C2DC5 was selected to identify the detailed mineralogy associated with the nanoSIMS derived isotope maps. Two ROIs from clast C2DC5 (see Data S2) were carefully extracted by focused ion beam (FIB) in situ lift-out technique with an FEI Helios SEM, equipped with an Omniprobe manipulator (WEMS lab, Cambridge University). Before the lift-out, ROIs were covered by a platinum organometallic complex $(C_9H_{16}Pt)$ of 1.5 (h) × 2 (w) × 15 (l) μm to protect the sample from unwanted damage due to a Ga ion beam, using first an electron beam and subsequently a $Ga^+$ beam to make a denser PtC protection layer. The section was then freed by stepwise ion-milling and attached to a Cu TEM half-grid, where it was thinned to electron transparency (∼100 nm) using a 30 kV Ga ion beam at currents reduced from 1 nA to 10 pA as the sample became thinner and a 2 kV Ga ion beam for final polishing. During the FIB lift-out procedure, the preparation advancement of the thin sections was occasionally observed using an electron beam with an estimated total dose of $10^3$-$10^4$ e/nm². The samples were then analyzed by an FEI Tecnai OSIRIS TEM at 200 keV (WEMS lab, Cambridge University), where Bright Field (BF) images were acquired in TEM mode, and High-Angle Annular Dark Field (HAADF) images and EDX maps were taken in STEM mode.

Electron Energy-Loss Spectra (EELS) were carried out on the Probe corrected Thermo Fisher Scientific Spectra 300 (WEMS lab, Cambridge University) with Gatan Continuum EELS spectrometer. The system was operated in energy filtered STEM mode, operating at 120 kV, convergence angle 0.764 mrad and a beam current of 100 pA, resulting probe size of 4.5 nm. We obtained spectral maps of organic-rich regions over a range of 245-520 eV using a sub-pixel scan, with a step size of 30 nm and a 2 sec integration time. A monochromator was used to obtain a high energy resolution of 0.1–0.2 eV, which provided distinct peak separations in carbon-K edges.

Sample excavation, digestion, purification, and multi-collector ICPMS (MC-ICPMS)

Four dark clasts were selected for further excavation, digestion, and isotope analyses (Table S4 and S5). Three clasts are large (A1DC1, A2DC1 and A3DC1) and one is small (C2DC5). For all clasts, we have corresponding SEM and LA-ICPMS data. For the A3 and C2 clasts, we also have nanoSIMS data and for C2 clast we have TEM data. Note that clast A2 has been heated, but the original clast structure is still visible. All clasts have been extracted using a computer assisted NewWave micro-drill with tungsten carbide drill bits at the Centre for Star and Planet formation (Copenhagen). After excavation, all clasts were carefully examined by high-resolution optical microscopy to verify that the samples were not contaminated by the surrounding matrix of the host meteorite. Note that the dark clasts are visually very distinctive from the host. The obtained powders were transferred in MQ water to clean Savillex beakers and were taken up in a $HNO_3$. HF mixture to be digested in Parr bombs for two days at 210°C. Afterwards, the samples were dried down and were further digested on a hotplate at 130°C in aqua regia. A 5% aliquot of a final 250 μl HCl solution was taken for iCAP inductively coupled plasma mass spectrometry (ICPMS) analyses to determine $^{27}Al/^{24}Mg$ and $^{55}Mn/^{52}Cr$ ratios. The HCl solution was then loaded on a 200-400 mesh Eichrom AG1-X8 anion resin (250 μl) to separate Fe from the matrix. This step was repeated to ensure complete removal of Ni from the Fe cut. The matrix cut was pretreated in a $1M\ HNO_3 + 10\%\ H_2O_2$ solution for 5 days and then loaded on a Dionex micro-cation column (200 μl AG50X12) where Mg and Cr cuts were collected with increasing $HNO_3$ concentration and flow rate over the column. Subsequently, Mg cuts were loaded on a Dionex CS16 column, where Mg was separated from Ni. Finally, the Cr cut was subjected to one or more Fe clean-up steps using a small anion column. Fe (10, 80, 200, 300 μg), Mg (5, 50, 100, 150 μg) and Cr (100 ng, 2, 2, 8 μg) cuts were fluxed in aqua regia for 2 days before being analyzed by Neptune Plus MC-ICPMS. Clast C2 was too small for Cr isotope analyses.

The Fe and Cr isotope compositions were measured using the Thermo Fisher Pandora Neptune Plus MC-ICP-MS at the Centre for Star and Planet formation in the high-resolution mode (M/ΔM = 12000 as defined by the peak edge width from 5 to 95% full peak height) to resolve gas-based interferences on the high mass side. Fe aliquots were introduced to the plasma source using an ESI Apex IF sample introduction system with an ACM unit attached and with an uptake of 30 µl min$^{-1}$. The detailed iron isotope acquisition setup follows Schiller et al. (*48*), where $^{60}$Ni and $^{53}$Cr were measured along $^{54}$Fe, $^{56}$Fe, $^{57}$Fe and $^{58}$Fe to correct for direct isobaric interferences on $^{54}$Fe and $^{58}$Fe from $^{54}$Cr and $^{58}$Ni, respectively. We report our data relative to the IRMM-014 Fe isotope standard in the µ-notation, where the reported data represent the mean and 2SE of 5-10 individual standard-bracketed sample analyses, each comprising 25 ×16.7 s of on-peak baseline measurement and 200 × 8.3 s of sample measurement. Cr aliquots were introduced using an ESI Apex HF sample introduction system with ACM unit and an uptake rate of 80 µl min$^{-1}$. The samples were measured without the use of an auxiliary gas to the introduction system to reduce gas-based interferences, at low RF power and sample gas inflow, with a Jet and X cone at medium resolution (M/ΔM > 5000). The Cr isotope setup otherwise follows Schiller et al. (*73*), where $^{49}$Ti, $^{51}$V and $^{56}$Fe were analyzed alongside $^{50}$Cr, $^{52}$Cr, $^{53}$Cr and $^{54}$Cr to correct for isobaric interferences. The measured interferences of $^{50}$V, $^{50}$Ti and $^{54}$Fe were 0.6 and 4.6 ppm, 850 and 1500 ppm and 2900 and 6500 ppm for clast A3 and A2, respectively. Doping tests of Ti, V and Fe in Cr solutions show that such interferences have a negligible effect on the µ$^{54}$Cr values (*74*). The largest Fe interference of 6500 ppm is estimated to decrease the µ$^{54}$Cr value by 10 ppm (*74*), within error of the analyzed A2 clast. Data is reported relative to the SRM979 Cr isotope standard in the µ-notation, where the reported data represent the mean and 2SE of 5 individual standard-bracketed sample analyses, each comprising 75 s of on-peak baseline measurement and 100 × 8.3 s of sample measurement.

Mg isotope compositions were measured using the Thermo Fisher Prometheus Neptune Plus MC-ICP-MS at the Centre for Star and Planet formation in the medium-resolution mode (M/ΔM > 5000) using an ESI Apex IR introduction system with ACM unit and an uptake rate of 30 µl min$^{-1}$. The analytical setup was done following Bizzarro et al. (*75*), where $^{24}$Mg, $^{25}$Mg and $^{26}$Mg were analyzed. Mg isotope data is reported relative to the DTS-2b standard in the µ-notation, where the reported data represent the mean and 2SE of 5-10 individual standard-bracketed sample analyses, each comprising 25 × 16.7 s of on-peak baseline measurement and 100 × 16.7 s of sample measurement.

Along with the dark clasts, we processed seven CR chondrules and a powdered Tarda aliquot of 200 mg through the same digestion and column chemistry to validate the analytical techniques used here, since CR chondrules constrain tight isotope distributions for Fe, Cr and Mg (*48, 76*) and have similar sizes and major element compositions as the dark clasts, whereas Tarda has a similar petrological make-up as the dark clasts. The average Fe, Cr and Mg isotope composition of these chondrules is µ$^{54}$Fe = 37±16 ppm, µ$^{26}$Mg* = −6.1±5.0 ppm and µ$^{54}$Cr = 121±33 ppm (2SD), in complete agreement with literature data (*48, 76*). The µ$^{54}$Fe and µ$^{26}$Mg*value for Tarda is 23.9±2.5 ppm and 2.7±2.1 ppm (2SE), respectively, in agreement with literature values (*77, 78*).

Three dark clasts have been analyzed for Si isotope compositions: A1DC1, A3DC1 and A3DC2 (Table S5). The chemical separation and mass spectrometric techniques used to acquire the Si isotope data have been described in detail in a previous study (*49*), but a summary below outlines the main features of the procedure. Sample powders were digested by NaOH fusion in Ag crucibles at 720 °C for 13 min following the protocol developed by (*79*). The resultant fusion cake was dissolved in 18.2 MΩ cm$^{-1}$ water and acidified with HNO$_3$. Following dissolution of the samples, purification of silicon was achieved by cation exchange chromatography (*79*). Columns were filled with 3 ml BioRad DOWEX 50W-X12 (200–400 mesh) resin in H$^+$ form. The total amount of silicon loaded onto the columns was ~150 µg for samples and ~300 µg for quartz-sand standard NBS-28. To minimize matrix effects arising from SO4$^{2-}$ ions not retained by the cation exchange resin, the solutions were doped with a sulfate solution to achieve a fixed S/Si ratio of ~5, following the approach of (*80*). Silicon isotope measurements were conducted using a Thermo Scientific Neptune Plus Multi-Collector Inductively Coupled Plasma Mass Spectrometer (MC-ICP-MS) at the Centre for Star and Planet Formation, University of Copenhagen, operating in high-resolution mode (M/ΔM ≈ 11,000). This configuration enabled the resolution of significant polyatomic interferences, notably $^{28}$Si$^1$H$^+$ (28.984762 amu) on $^{29}$Si (28.97649 amu) and $^{14}$N$^{16}$O$^+$ (29.99799 amu) on $^{30}$Si (29.97377 amu). Sample solutions with Si concentrations of approximately 2-3 ppm were introduced into the plasma source in 0.5 M HNO$_3$ via an Apex-Q system equipped with an ACM Nafion fluoropolymer membrane desolvation module, at an uptake rate of approximately 100 µL/min. The instrumental sensitivity at this uptake rate was typically 20 V ppm$^{-1}$, and measurements were performed at ~45 V on $^{28}$Si. Each sample underwent ten analyses in a single session, with each analysis comprising 100 measurements lasting ~8.4 seconds. On peak blank measurements, accounting for less than 0.1% of the total signal, were bracketed for each sample and standard and subtracted. Samples were internally normalized to a $^{29}$Si/$^{28}$Si ratio of 0.0508 and are presented as µ units relative to the quartz-sand standard NBS-28 (NIST RM8546), serving as the bracketing standard for external normalization. Along with the samples, a BHVO-2 standard was processed and analyzed, returning a $\mu^{30}Si$ value of -0.5±7.9 ppm (2SE). The precision of Si isotope measurements aligns with the external reproducibility of our method, as detailed in (*49*).

**Acknowledgments:** We thank Michael Küffmeier and Troels Haugbølle for their discussion on various aspects of this work. We appreciate the helpful contributions of three anonymous reviewers that have improved our manuscript.

**Funding:**

Funding for this project was provided by the Villum Foundation (Villum Young Investigator Grant #53024) to E.v.K as well as the Carlsberg Foundation (CF18_1105) and the European Research Council (ERC Advanced grant Agreement 833275—DEEPTIME) to M.B. M.B. acknowledges support from the Villum Foundation (Villum Investigator Grant #54476). M.S. acknowledges support from the Carlsberg Foundation (CF20_0209) and the Villum Foundation (00025333).

European Union's Horizon 2020 research and innovation programme under grant agreement No 101005611 for Transnational Access conducted at the Wolfson Electron Microscope Suite, University of Cambridge, Department of Materials Science and Metallurgy (EvK). We acknowledge use of the Thermo Fisher Spectra 300 TEM funded by EPSRC under grant EP/R008779/1.

**Author contributions:**

Conceptualization: EvK, MB, MS

Methodology: EvK, MB, MS

Investigation: EvK, IF, XZ, JW, PT, SF, IO, MS

Writing – original draft: EvK, MS

Writing – review & editing: EvK, MB, MS, IF

Resources: MB, JW, PT, XZ, EvK

Funding acquisition: EvK, MB, MS

Visualization: EvK, MB

Supervision: JW, EvK

Data curation: PT, EvK

Validation: PT, IO, IF, XZ, EvK

Formal analysis: EvK, XZ

Project administration: EvK

**Competing interests:** The authors declare that they have no competing interests.

**Data and materials availability:** All data are available in the main text or the supplementary materials.

**Supplementary Materials**

Supplementary Text

Figs. S1 to S16

Tables S1 to S5

References (*91–104*)

Data S1 and S2

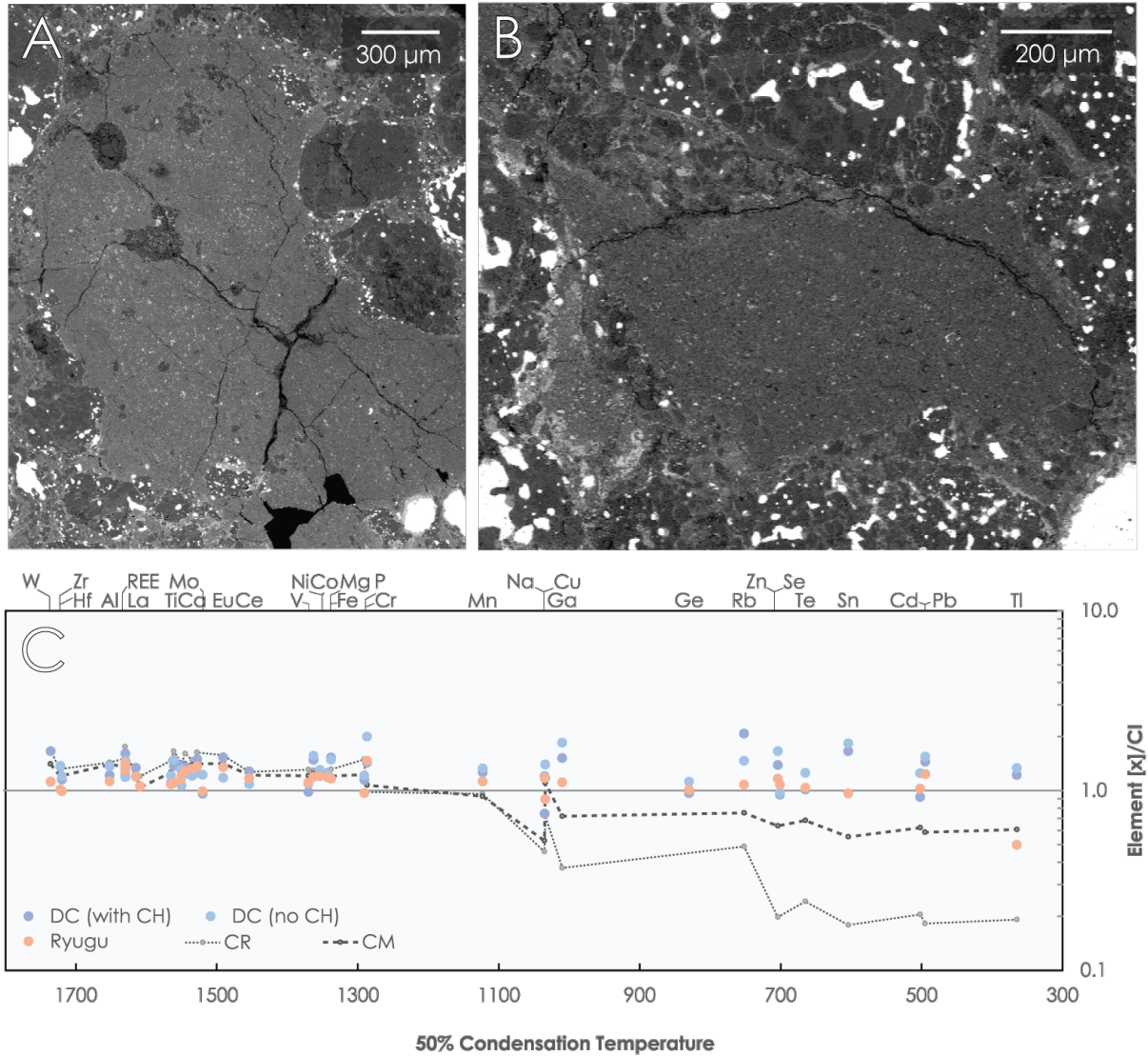

**Fig. 1. Petrology and composition of dark clasts from CR NWA 14250.** Back scattered electron (BSE) images from **A)** a large chondrule-containing clast and **B)** a smaller clast without chondrules. **C)** Averaged laser ablation inductively coupled plasma mass spectrometer analyses of the dark clast matrix (see Methods) show that the matrices of both clast types are chemically identical and resemble CI chondrites (y=1) (*81*) and Ryugu (orange) (*82*). Bulk CR and CM chondrite compositions are plotted for reference (*83*). 50% condensation temperatures are from (*84*). Additional BSE images, elemental maps, and compositional data are reported in the SM (Fig. S2-7) and Data S1. DC: dark clast. CH: chondrule.

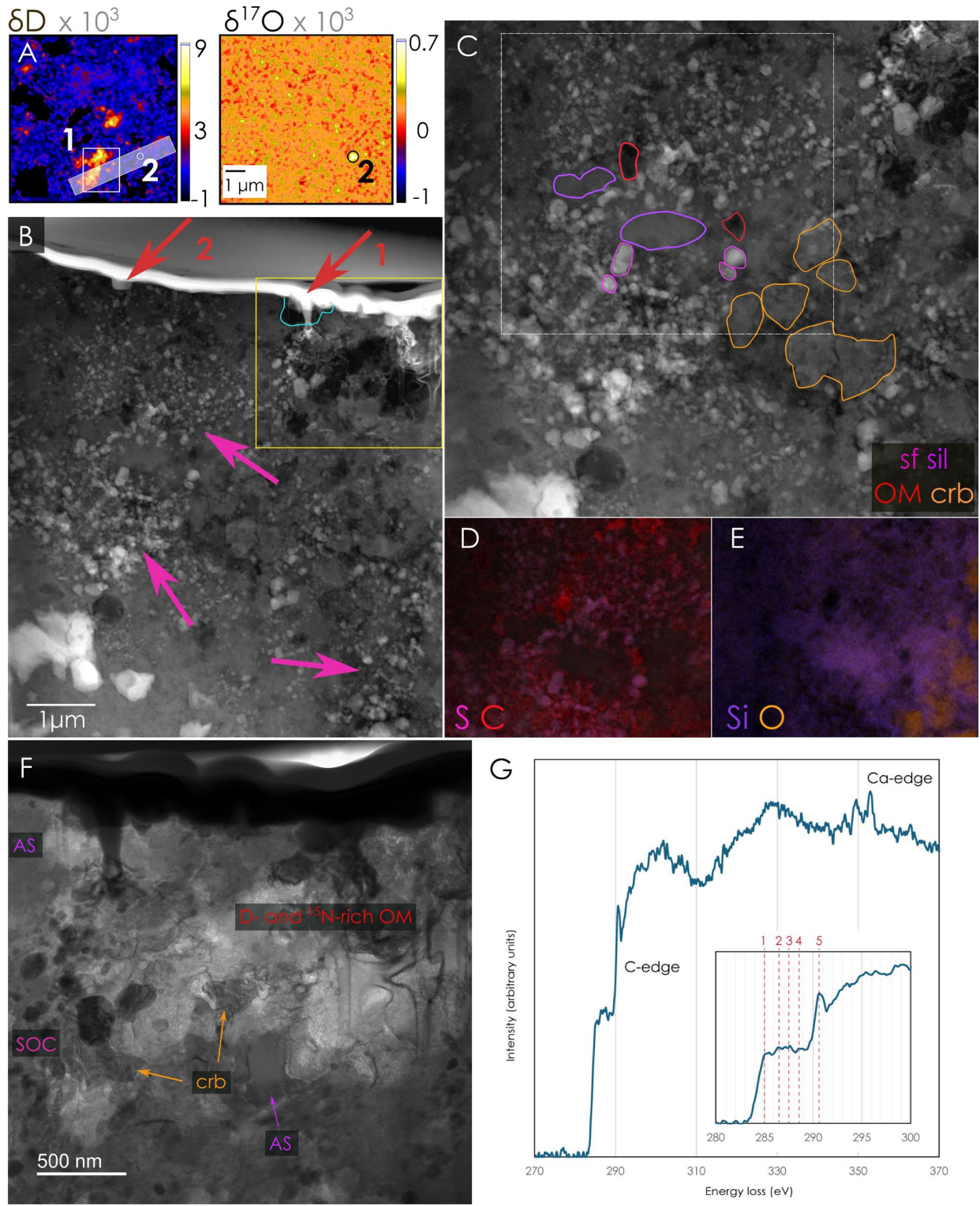

**Fig. 2**. **Transmission electron microscopy (TEM) analyses of dark clast matrix. A)** Nanoscale correlated hydrogen and oxygen isotope maps from dark clast C2DC5 (map 3), which includes a presolar oxygen-anomalous grain (region of interest: ROI#2) surrounded by D- and $^{15}$N-rich areas (ROI#1). The shaded white area shows the platinum protective strip placed before extraction of a matrix slice (**B**) High-angle annular dark field image including ROIs (red arrows). The pink arrows point to the abundant sulfide-organic-complexes (SOC). Yellow box depicts region shown in panel F and G with ROI#1 (D- and $^{15}$N-

rich organic matter). **C**) Zoom-in of panel B showing a white box where elemental mapping was carried out (panel **D** and **E**). Highlighted color-coded areas include sulfides (pink), carbonates (orange), organic matter (red) and amorphous silicates (purple). **F**) Zoomed in region of yellow box in panel A in BF mode, where white areas are organic-rich, grey is amorphous silicate (AS) or carbonate (crb) and black is sulfides. SOC are directly adjacent to D- and $^{15}$N-rich organic matter. **G**) Electron Energy Loss Spectrum of organic matter in ROI#1 (panel F) showing the bonding environments of carbon (C-K edge) and calcium (Ca-K edge). Peaks in the Ca-K edge confirm the carbonate related bonding environments in the C-K edge. The inset shows the detailed C-K edge with peaks at 1) ~285 eV: aromatic bonds, and more oxygenated carbon at peaks: 2) ~286.5 eV: carbonyl ester and ether groups in aldehydes and ketones, nitrile bonding environments, 3) ~287 eV: aliphatic bonds, 4) ~288.5 eV: carbonyl groups in carboxyl moieties and 5) ~290.3 eV: carbonate bonds. Similar organic matter with high oxygenated over aromatic carbon structures, including carbonate associated with the organics, has been previously identified in cometary interplanetary dust particles (*26*). Additional TEM images are available in Fig. S8-S11.

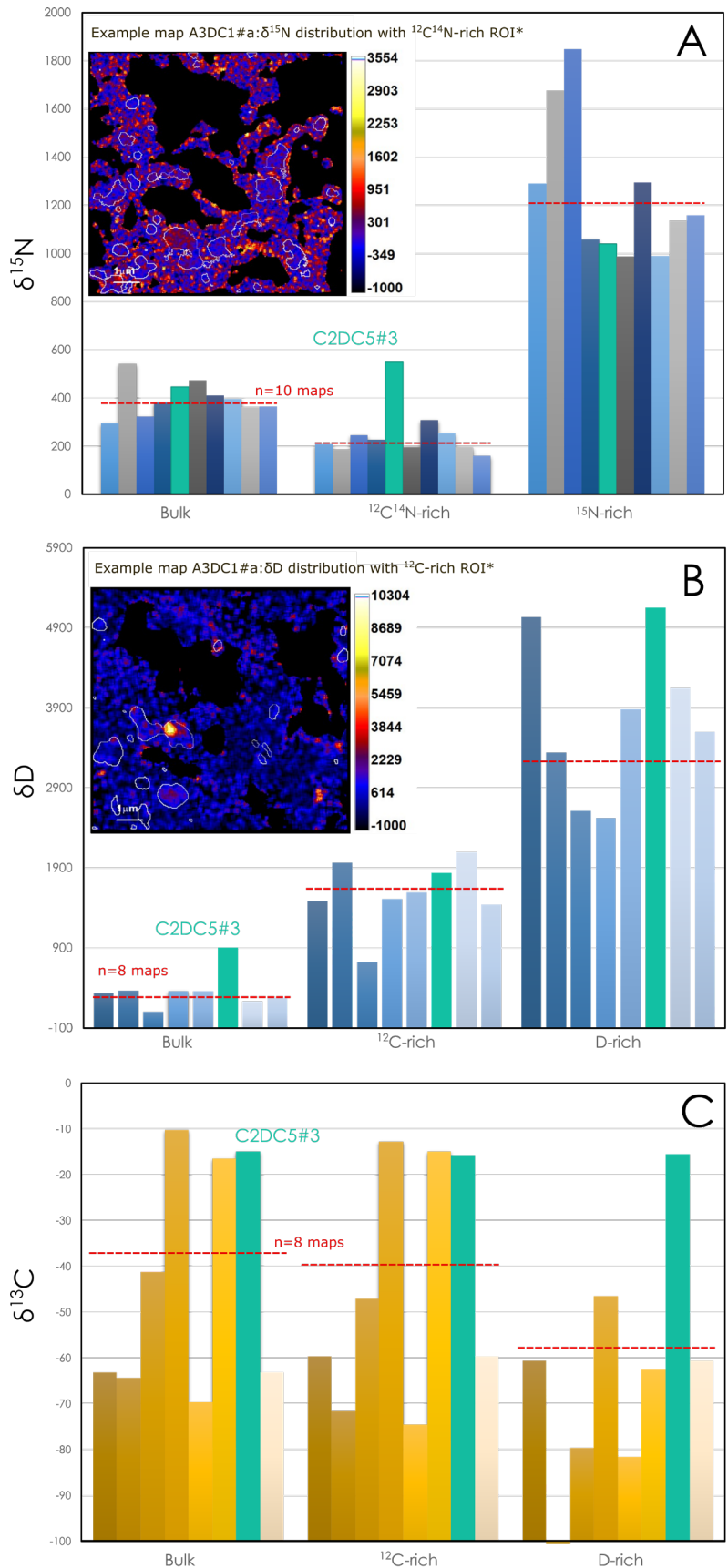

**Fig. 3: Nitrogen, hydrogen, and carbon isotope data of dark clasts.** The distribution of **A)** $\delta^{15}N$, **B)** $\delta D$ and **C)** $\delta^{13}C$ in the bulk clasts, the carbon-rich areas and the $^{15}$N- and D-rich areas from the least altered matrix locations from A3DC1, A3DC2 and C2DC5. Individual maps, raw data and data reduction methods can be found in Data S2. *The $^{12}C^{14}N$ and $^{12}$C-rich areas are determined using a 20% counts cut-off from the lower limit. The $^{15}$N-rich hotspots are defined by the 20% most enriched pixels and the D-rich areas from the 2% most enriched pixels. Each bar represents an individual map. The hydrogen and carbon isotope maps are overlayed on the nitrogen isotope maps but note that not all nitrogen isotope maps have a corresponding hydrogen and carbon isotope map. Additional maps from more altered areas (i.e. clast C3DC4) can be found in the supplementary materials. The green bars represent data from map C2DC5#3 that contains an oxygen-anomalous presolar grain and accompanying anomalous D-rich organic matter.

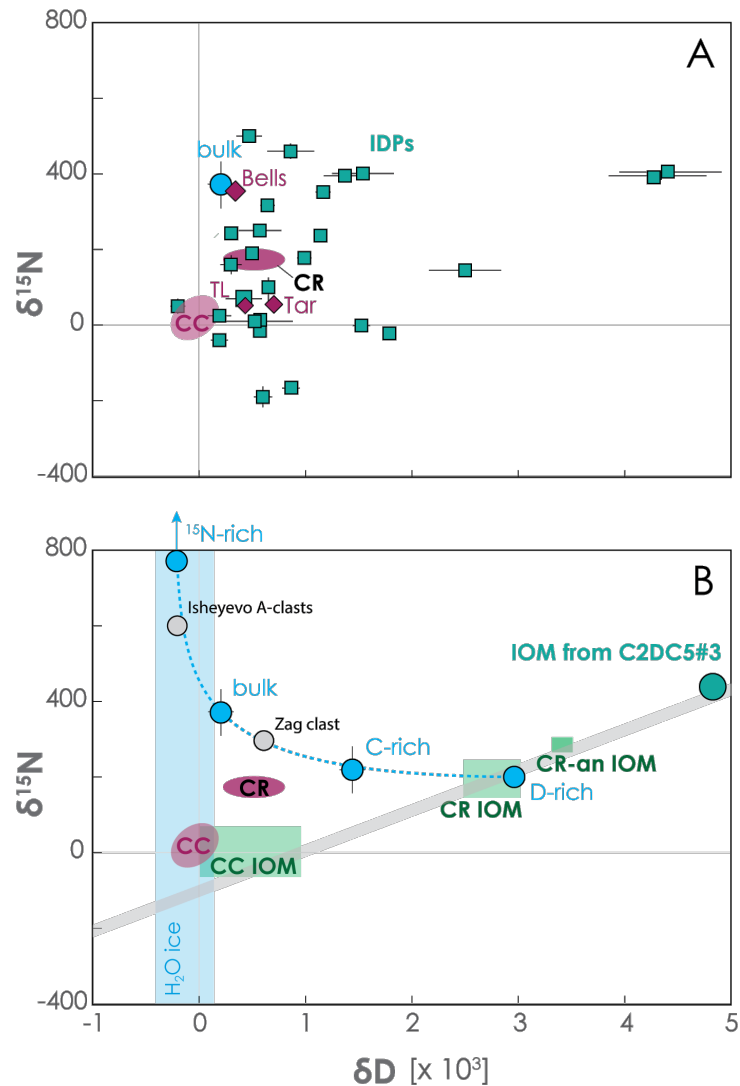

**Fig. 4: δ¹⁵N versus δD values of Solar System materials. A)** Bulk carbonaceous chondrites (CC: purple fields and diamonds, TL = Tagish Lake, Tar = Tarda) (*85*) and interplanetary dust particles (IDPs: squares) (*26*, *44*, *86*), with the average bulk composition of dark clasts (from Fig. 3). **B)** Same plot as panel A but with dark clast data from C-rich and D-rich regions (from Fig. 3). Grey bar: evolution of organic matter through cold ion-molecule reactions in the interstellar medium (*30*), blue bar: the range of initial δD values for water ice from CC groups and dark clasts (*31*, *33*). We also plot insoluble organic matter (IOM) of CCs and CR chondrites (*56*) and the IOM from the oxygen-anomalous presolar grain area in map C2DC5#3 (see Fig. 2). A blue mixing line plots through the average bulk dark clasts, the average ¹²C-rich areas, and the D-rich endmember. The other endmember is ¹⁵N-rich and D-poor (δD value taken from Fig. S13). See SM for mixing line calculations (Table S2 and Fig. S14). Note that if we include available data for the Zag clast (*11*, *12*) and anhydrous lithic clasts from Isheyevo (*31*) with proposed outer disk origins, these data fall on the same mixing line. Other bulk xenolithic clasts with correlated δD and δ¹⁵N data (*13*, *17*) show closer affiliations to carbonaceous chondrites and are not plotted for this reason. Observe that the nitrogen and hydrogen isotope measurements of IDPs and the clasts were obtained by in situ techniques whereas bulk carbonaceous chondrites and their IOM were analyzed by gas spectrometry. However, potential analytical biases between these techniques are comparable to the uncertainties of our measurement (*33*, *54*, *87*) and are, hence, negligible.

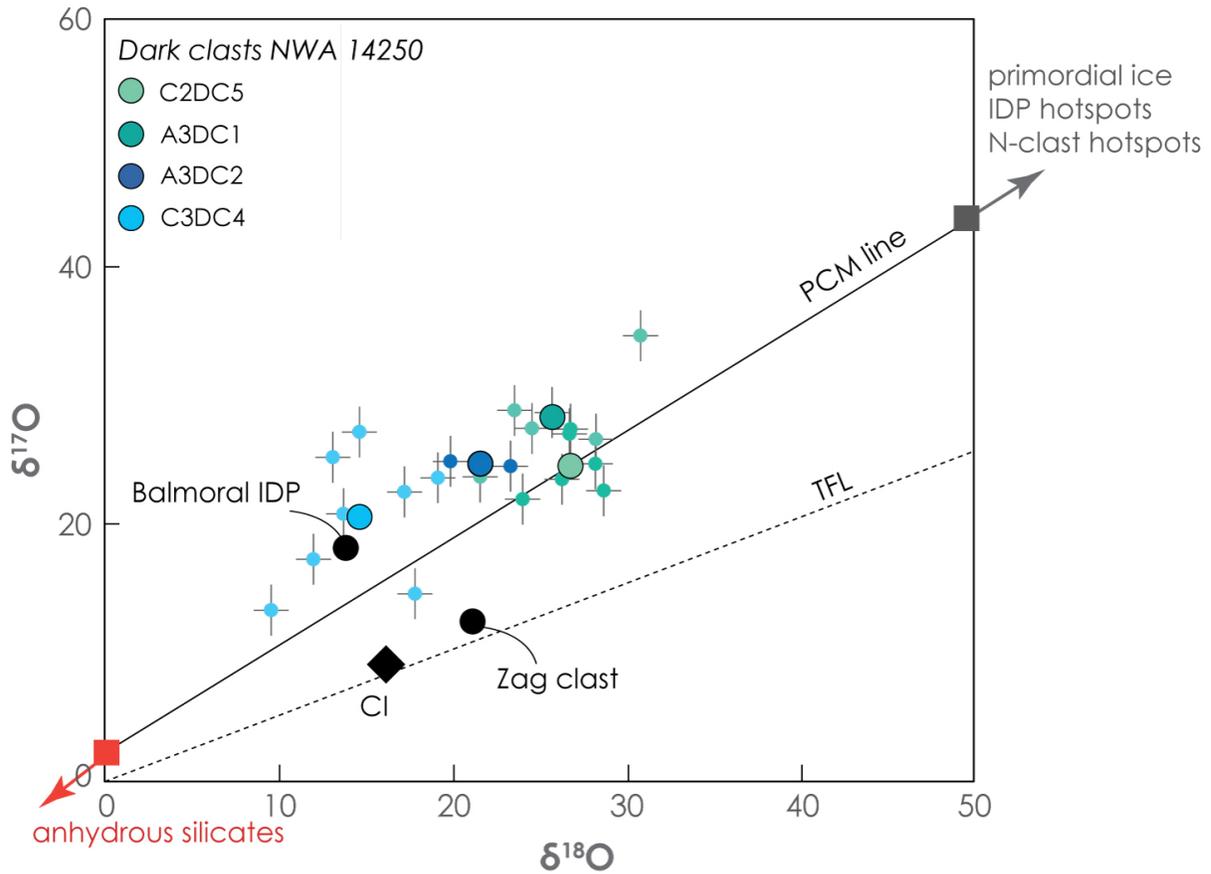

**Fig. 5: Three oxygen isotope plot** showing the terrestrial fractionation line (TFL) and slope 1 mixing line (PCM = Primordial Component Mixing) (*88*) between a $^{16}$O-poor and a $^{16}$O-rich endmember produced by self-shielding (*38*, *42*). The $^{16}$O-poor endmember reflects the primordial interstellar medium (ISM) ice composition inferred from cosmic symplectites in Acfer 094 (*89*), hotspots from interplanetary dust particles (IDPs) (*3*) and $^{16}$O-poor hotspots from the potentially cometary clast (N-clast after (*17*)) found in the CR chondrite LAP 02342. The bulk oxygen isotope maps of the dark clasts (small colored spheres) fall along this line and are the most $^{16}$O-depleted bulk Solar System materials found so far. Large colored spheres represent averaged maps of clasts. We show the bulk CI chondrite composition (*52*), the Zag clast (*12*) and the Balmoral IDP (*44*).

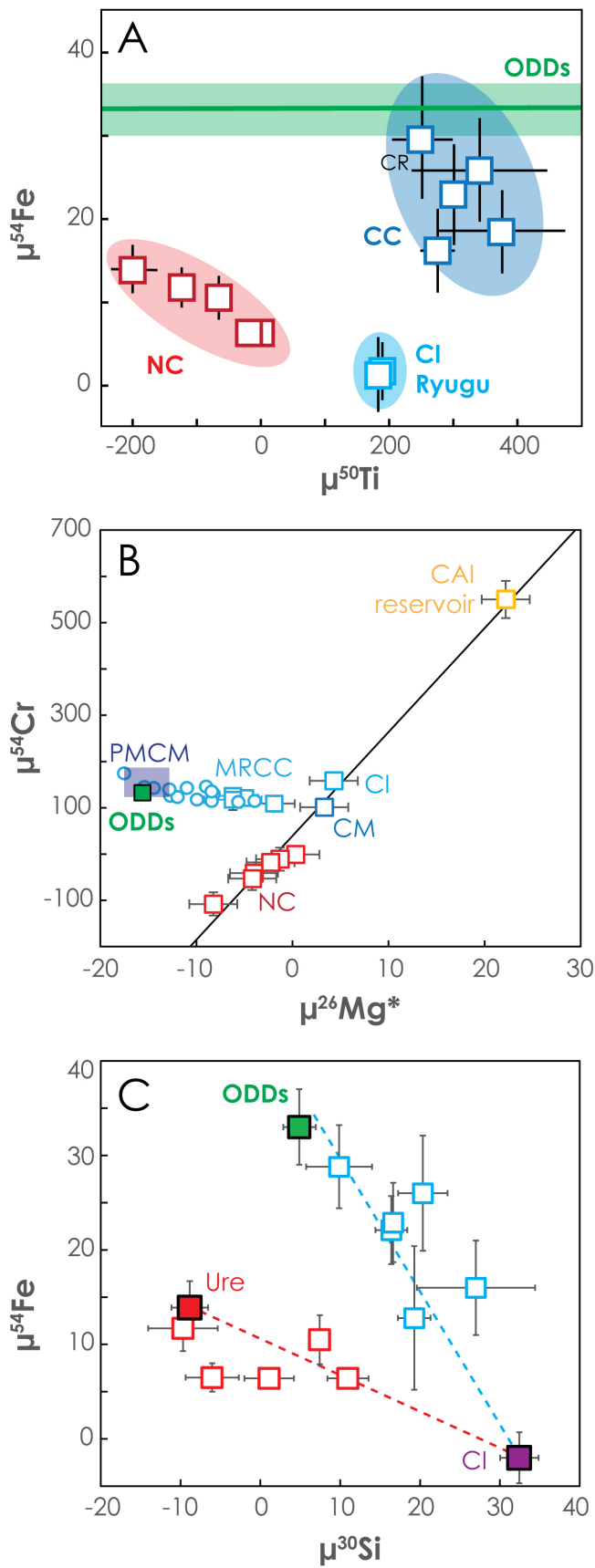

**Fig 6: The nucleosynthetic fingerprint of the comet-forming region. A**) The μ⁵⁴Fe value of outer disk dark clasts (ODDs) plotted together with non-carbonaceous (NC) and carbonaceous (CC) chondrite as well as CI and Ryugu. **B**) μ²⁶Mg*-μ⁵⁴Cr systematics of ODDs show that their composition is not similar to calcium-aluminum-inclusions (CAIs) but akin to that predicted for ²⁶Al-free Primordial Molecular Cloud Matter (PMCM). **C**) The Solar System trichotomy, consisting of three endmember compositions in μ³⁰Si-μ⁵⁴Fe space. Ure = ureilite. Note that panel A reflects a similar trichotomy, but CC are shifted to higher μ⁵⁰Ti values due to addition of CAIs (*7*).

.

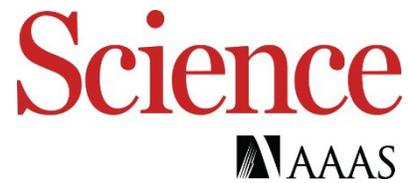

Supplementary Materials for

**The nucleosynthetic fingerprint of the outermost protoplanetary disk and early Solar System dynamics**

Elishevah van Kooten, Xuchao Zhao, Ian Franchi, Po-Yen Tung, Simon Fairclough, John Walmsley, Isaac Onyett, Martin Schiller, Martin Bizzarro

Corresponding author: elishevah.vankooten@sund.ku.dk

**The PDF file includes:**

> Supplementary Text
> Figs. S1 to S16
> Tables S1 to S5
> References

**Other Supplementary Materials for this manuscript include the following:**

> Data S1 to S2

## Supplementary Text

<u>The degree of secondary alteration on CR NWA 14250</u>

We have analyzed 15 ferrous olivine grains from 4 type II chondrules in the recently classified CR chondrite NWA 14250 by LA-ICPMS at Lund University (Table S1). The mean $Cr_2O_3$ concentration of these grains is 0.41±0.08 wt.%, which we plot alongside other CR chondrite data1 (Fig. S1). We show that NWA 14250 has experienced very low degrees of thermal metamorphism based on the $Cr_2O_3$ content of ferrous olivine grains, since even minor heating will result in Cr loss from olivine (*19*). These measurements make clear that NWA 14250 is amongst the most pristine CR chondrites.

<u>NanoSIMS data</u>

Within the N isotope maps, we find two different distributions of $^{15}N$ and both types can be found in all analyzed clasts. In the first type (Fig. 3A and S12A), the bulk carbon-rich material is depleted in $^{15}N$ relative to the bulk and contains $^{15}N$-rich phases bordering the organic matter ($\delta^{15}N$ hotspots >1000 ‰). The second type (Fig. S12B) has similar $\delta^{15}N$ bulk values but more $^{15}N$-rich organic (carbon-rich) areas containing relatively low $\delta^{15}N$ hotspots (<800 ‰). As we explain in the main text, the N isotope distribution in Fig. 3A and S12A reflects the primary unaltered isotope distribution of the matrix since these maps contain the most $^{15}N$-hotspots. Homogenization of $^{15}N$ during secondary alteration will produce $^{15}N$ distributions as shown in Figure S12B. Hence, the organic matter is initially relatively $^{15}N$-poor (~200 ‰) but becomes enriched in $^{15}N$ during diffusion (perhaps shock-induced) of $^{15}N$-rich outer layers. This process could explain why small organic globules (<200 nm) with a relatively high surface area become relatively more enriched in $^{15}N$ and appear as hotspots that correlate with D/H ratios (*90*).

<u>Mixing lines in δD versus $\delta^{15}N$ space</u>

We have calculated the ideal mixing line through the average bulk dark clast composition ($\delta^{15}N$ = 378±32 ‰, δD = 274±74 ‰), the average C-rich areas ($\delta^{15}N$ = 210±25 ‰, δD = 1462±92 ‰) and the D-rich organic matter (OM), with δD ≈ 3000 ‰ and $\delta^{15}N$ ≈ 200 ‰. The δD values are taken from Fig. S12C and the $\delta^{15}N$ values are from Fig. S12A. The D-rich organic matter was used as the first endmember and the H/N ratio (22:1) was calculated using the averaged IOM composition for CR chondrites (*56*) (Table S2). The second endmember reflects the fluid composition, which is assumed to be dominantly water with a δD of −206 ‰, as calculated in Fig. S13. The $\delta^{15}N$ value and the H/N ratio of this fluid are the only variables that are adjusted in these mixing line calculations to produce a mixing line that fits all three points (Fig. S14, Table S2). We find a best fit, when we add <1% of $NH_3$ ice to the fluid, with a H:N ratio of 5000. This $NH_3$ ice is very enriched in $^{15}N$, with a $\delta^{15}N$ value of ~6500 ‰. Mixing equations of two isotope endmembers are detailed in DePaolo and Wasserburg (*91*).

<u>Presolar grains in the dark clasts</u>

Based on the carbon and oxygen isotope distribution within the dark clast matrix, two presolar grains were identified, namely one SiC grain with a $\delta^{13}C$ value of 403±24 ‰ and one O-anomalous grain with $\delta^{17}O$ and $\delta^{18}O$ values of 539±66 ‰ and −214±24 ‰, respectively (see Data S2). We note that the region where the O-anomalous presolar grain is located contains the most anomalous D-rich (Fig. S12B) and $^{15}N$-rich (Fig. S12A) organic matter, which is highly enriched relative to chondritic IOM (i.e., insoluble organic matter, Fig. 4). The spatial association of the O-anomalous presolar grain with the D-rich and $^{15}N$-rich composition of the organic matter preserved in these dark clasts reinforces the idea of an ISM origin for these materials.

<u>TEM: Primary sulfide and Mg-rich carbonate in sulfide-organic complexes (SOC)</u>

Glassy Embedded Metal and Sulfide assemblages (GEMS) in chondritic-porous IDPs have been attributed an interstellar origin (*25*), based on the amorphous nature of the silicates and the presolar isotope compositions of some GEMS (*92, 93*). GEMS-like materials have also been observed in the most unaltered chondritic matrices, but it remains an open question whether these materials are related to IDP GEMS or not (*24, 94, 95*). The main consent is that GEMS-like materials have been attributed disequilibrium condensation origins from a high-temperature events in the solar nebula and are unrelated to GEMS in IDPs. The distinction between chondritic and IDP GEMS is made by the low Fe content in the amorphous silicate groundmass (<5 at.%) of IDP GEMS relative to that of chondritic GEMS (<25 at.%) (*24*). Other characteristic differences include the size of GEMS (typically <1 μm in IDPs and > 1 μm in chondritic matrix) and the wider range of sulfide size down to the nm scale in IDP GEMS. In general, the GEMS-like objects in the dark clasts studied here have more in common with chondritic GEMS such as observed in Paris and Acfer 094, owing to their larger irregular sizes and the absence of kamacite (IDP GEMS typically contain bot kamacite and low-Ni pyrrhotite), although alteration rims around the sulfides as observed for the Paris meteorite (*94*) are not observed for the dark clasts in this study. Nevertheless, some features of the GEMS-like objects from the dark clasts (named SOC in the paper do to their high organic content) do not allow for a high-temperature origin because: 1) the SOC contain organic-rich material between the sulfides and can thus not be defined solely as amorphous silicates, more like an amorphous mixture of organics and silicate. These organics are not expected to survive during high-temperature condensation processes and genetic relationships between complex organics and GEMS have also been observed in anhydrous IDPs (*28*). 2) The N and H isotope compositions of the dark clast matrix are interpreted as having an outer Solar System origin and the organic-rich matter has an isotope composition that plots on the ISM cold condensation trend. 3) The overall composition of the matrix is solar and the SOC make up a significant fraction of the matrix (~40 vol.%). Since chondritic GEMS-like material has a non-solar composition (including probably significant volatile loss), it is unlikely that the SOC reflect a high-T condensation origin.

Sulfides in aqueously altered chondrites are typically secondary alteration products as opposed to primary ISM or protoplanetary disk phases. However, the observation that SOC are physically associated with amorphous silicates establishes that they are of primary origin. Even though GEMS-like chondritic materials including sulfides have been suggested to form through high-temperature condensation processes, organosulfur complexes as observed here have also been interpreted as having an ISM origin. Indeed, an ISM origin for SOC is consistent with formation of refractory organosulfur complexes through gas-grain reactions (*96*) or S ion bombardment (*29*), which may reflect the precursor materials to the SOC observed in this study. In addition to sulfides, organics, and amorphous silicates, we also find a phase associated with these complexes that is Si-poor and Ca-Mg-C-O-rich (Fig. 2E, S8, S10 and S11), which is most likely carbonate. Carbonates in carbonaceous chondrites are the product of progressive aqueous alteration on the chondrite parent body and their presence in the dark clast matrix may be interpreted as such. However, it should be noted that carbonates found in hydrous IDPs have been linked to the 6.8 μm feature in ISM spectra, which is suggested to reflect carbonates that have condensed in the cold regions of the outer Solar System (*97–99*). In addition, the presence of Mg-carbonates and lack of phyllosilicates in Stardust samples has been interpreted as a cold condensation origin (*100*) for the carbonates. The fact that carbonates in the dark clast matrix are restricted to pristine regions containing ubiquitous amorphous silicates and organic matter supports the idea that these carbonates have been inherited form the ISM. Therefore, the bulk of the material preserved in the dark clasts is interpreted to represent unmodified ISM-inherited matter, consistent with an outer disk origin for these objects.

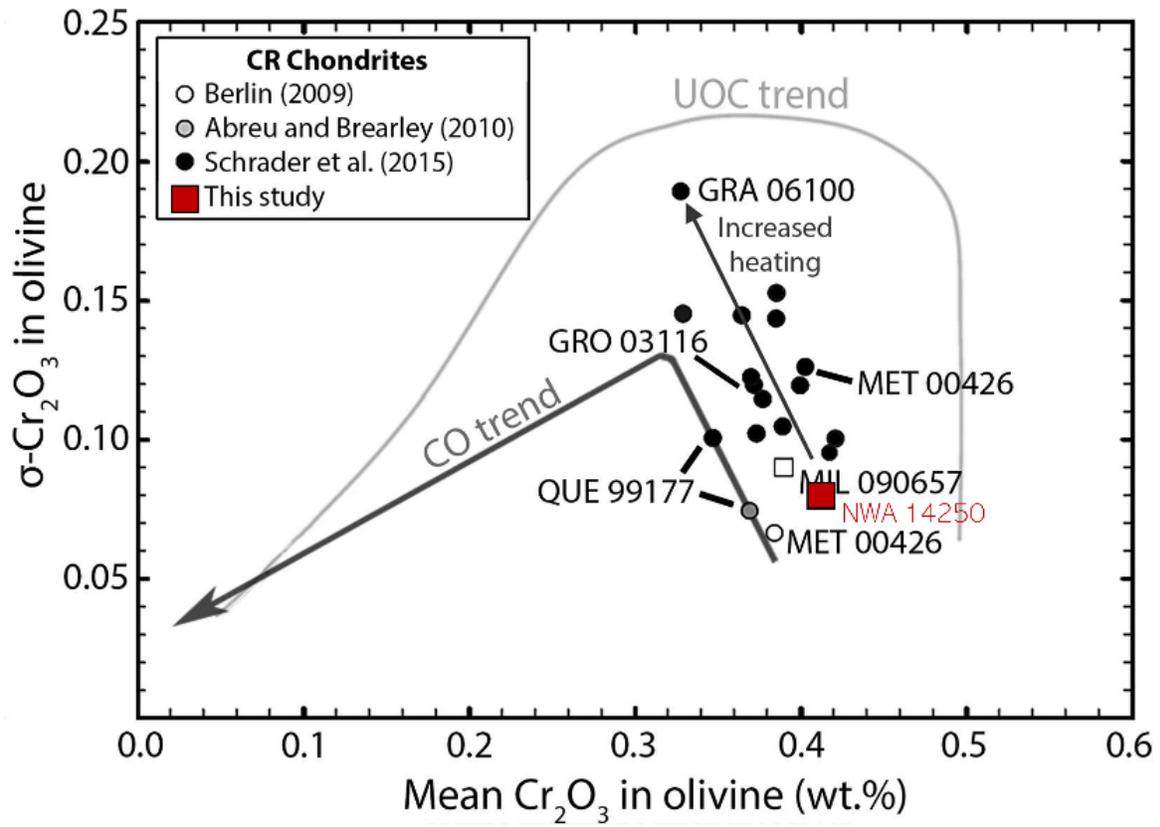

**Fig. S1:** Mean Cr$_2$O$_3$ content in olivine versus 1 sigma deviation from the mean. Plot modified after Davidson et al. (*19*)

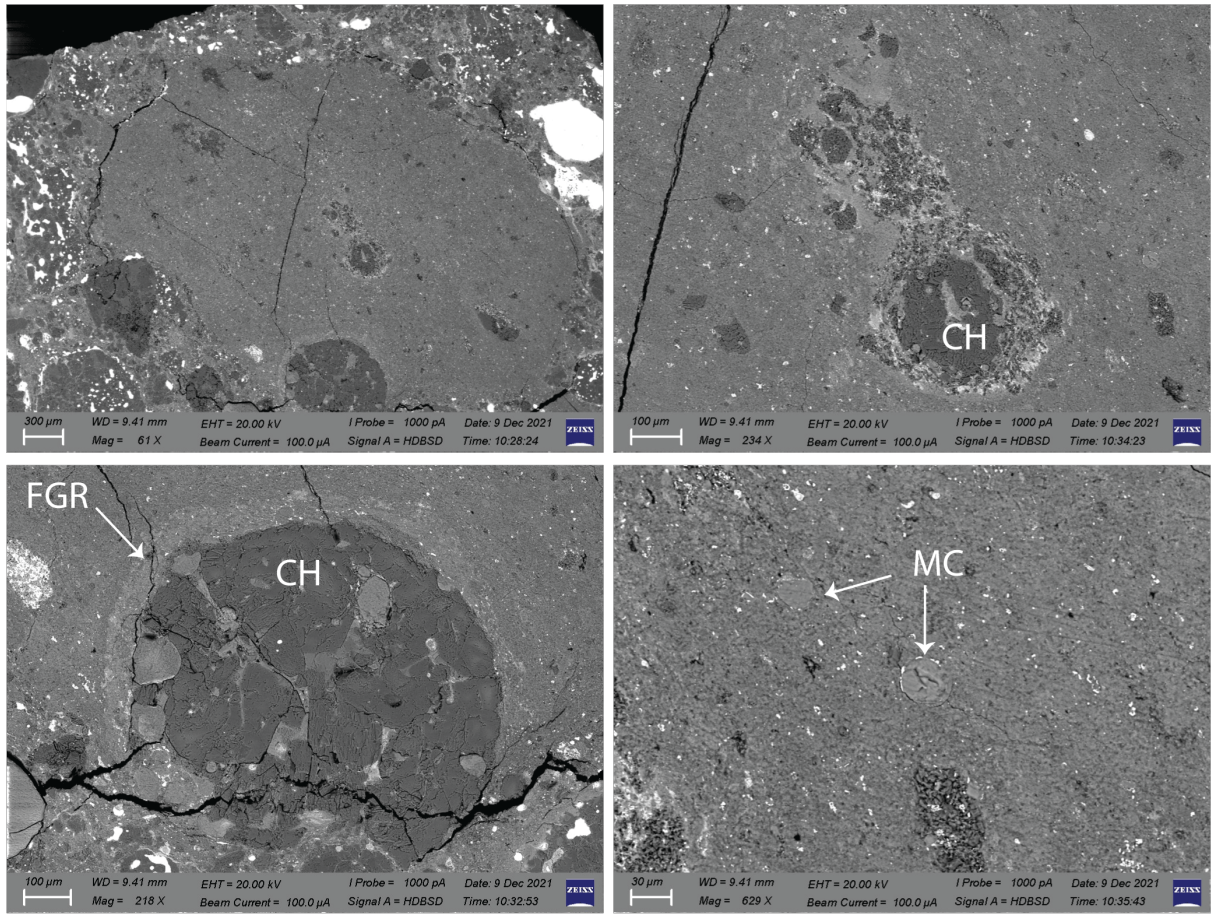

**Fig. S2: Back scattered electron images from dark clast A3DC1,** also used for nano secondary ion mass spectrometry analyses and sampled for Fe, Mg, Si and Cr isotope analyses (Table S4). Like the other large millimeter-sized clasts, <10 vol.% consists of chondrules (micro: <50 µm, median: ~200 µm and large: 300-500 µm; >200 µm typically contain fine-grained rims: FGRs) and fine-grained matrix with small sulfides as main accessory phase. MC = microchondrule, CH = chondrule. Although most chondrules in the studied clasts have close affinities to CM chondrules (i.e., size range of 100-200 µm, rare metal, fine-grained dust rims; Fig. S2-3), abundant glassy FeO-rich microchondrules (<50 µm, Fig. S4) and larger chondrules (<500 µm, Fig. S2) are also present. Chondrules in comet Wild2 returned by the Stardust mission are predominantly FeO-rich chondrules (type II) (*100*), which contrasts with the dominantly Fe-poor chondrules (type I) found in the dark clasts. However, the material returned by Stardust is biased towards small sized particles (<30 um) and, as such, may not be representative of the actual cometary chondrule inventory. Furthermore, comet Wild2 is depleted in carbon relative to mainstream comets (*100*). As carbon is a reducing agent, it has been suggested that the difference between reduced type I and oxidized type II chondrules reflects the variable availability of carbon in the protoplanetary disk (*4*). Hence, like chondrites, comets may contain a wide variety of chondrules.

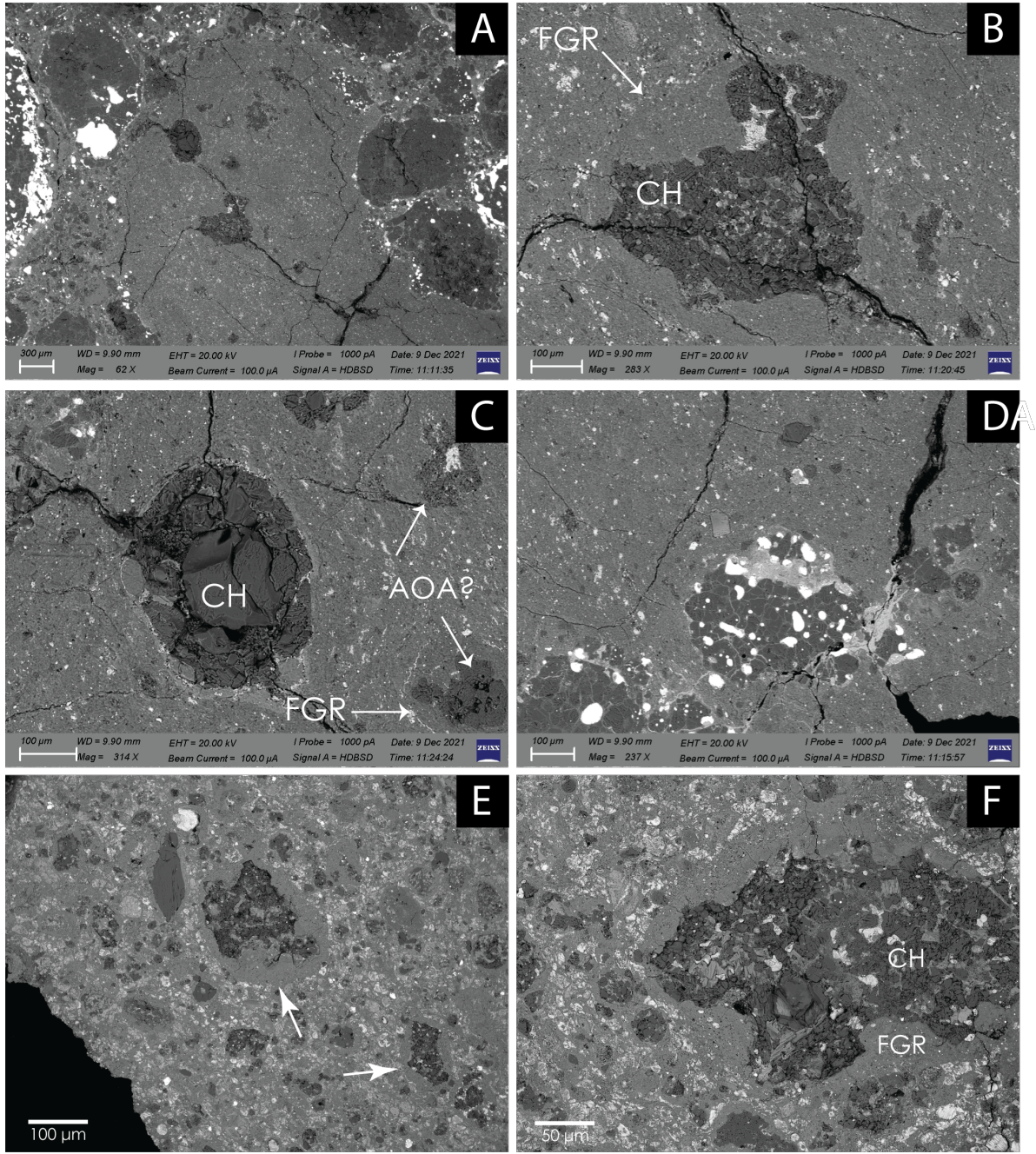

**Fig. S3: A-D) Back scattered electron images from dark clast A3DC2**, also selected for further nano secondary ion mass spectrometry and Si isotope analyses (Table S4). Like the other large mm-sized clasts, <10 vol.% consists of chondrules (micro: <50 µm, median: ~200 µm and large: 300-500 µm; >200 µm typically contain fine-grained rims: FGRs) and fine-grained matrix with small sulfides as main accessory phase. CH = chondrule. Some of the identified inclusions are potentially amoeboid olivine aggregates (AOAs), since they have an irregular shape and are fine-grained and porous in some cases. These objects are typically <100 µm in diameter. However, similar objects have also been observed in CM chondrites and are identified as chondrules. **E-F)** Representative chondrules from the CM chondrite Maribo, with the same irregular shapes, filled up by FGRs as found in the dark clasts. The porphyritic olivine chondrules are finely grained and the chondrules resemble AOAs to some extent. FGR: fine-grained rims, CH: chondrules, AOA: amoeboid olive aggregate.

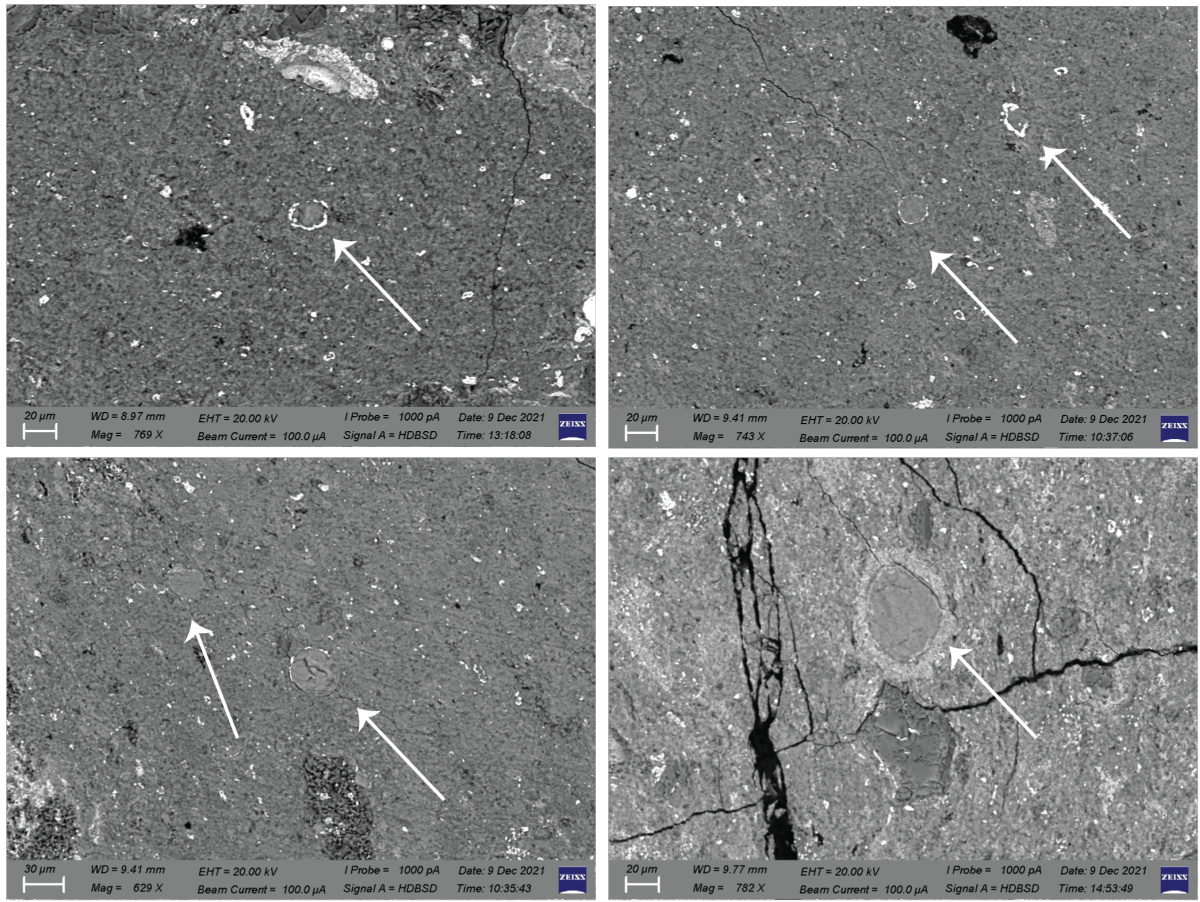

**Fig. S4: Back scattered electron images of microchondrules** from four different dark clasts in this study. Many microchondrules appear glassy and are surrounded by metal/sulfide rims.

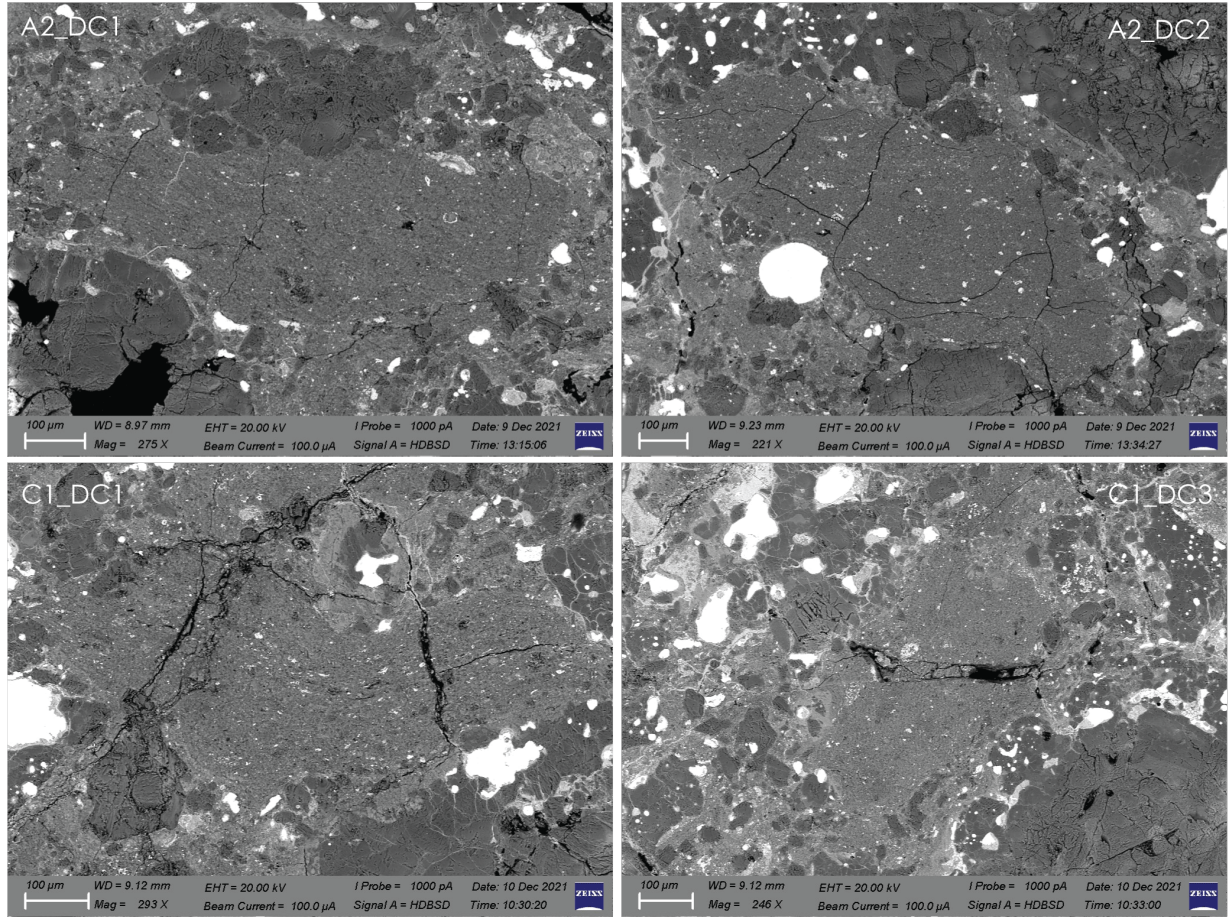

**Fig. S5:** Relatively small sub-millimeter sized dark clasts with only a few microchondrules as inclusions. These clasts likely represent fragments of the larger clasts.

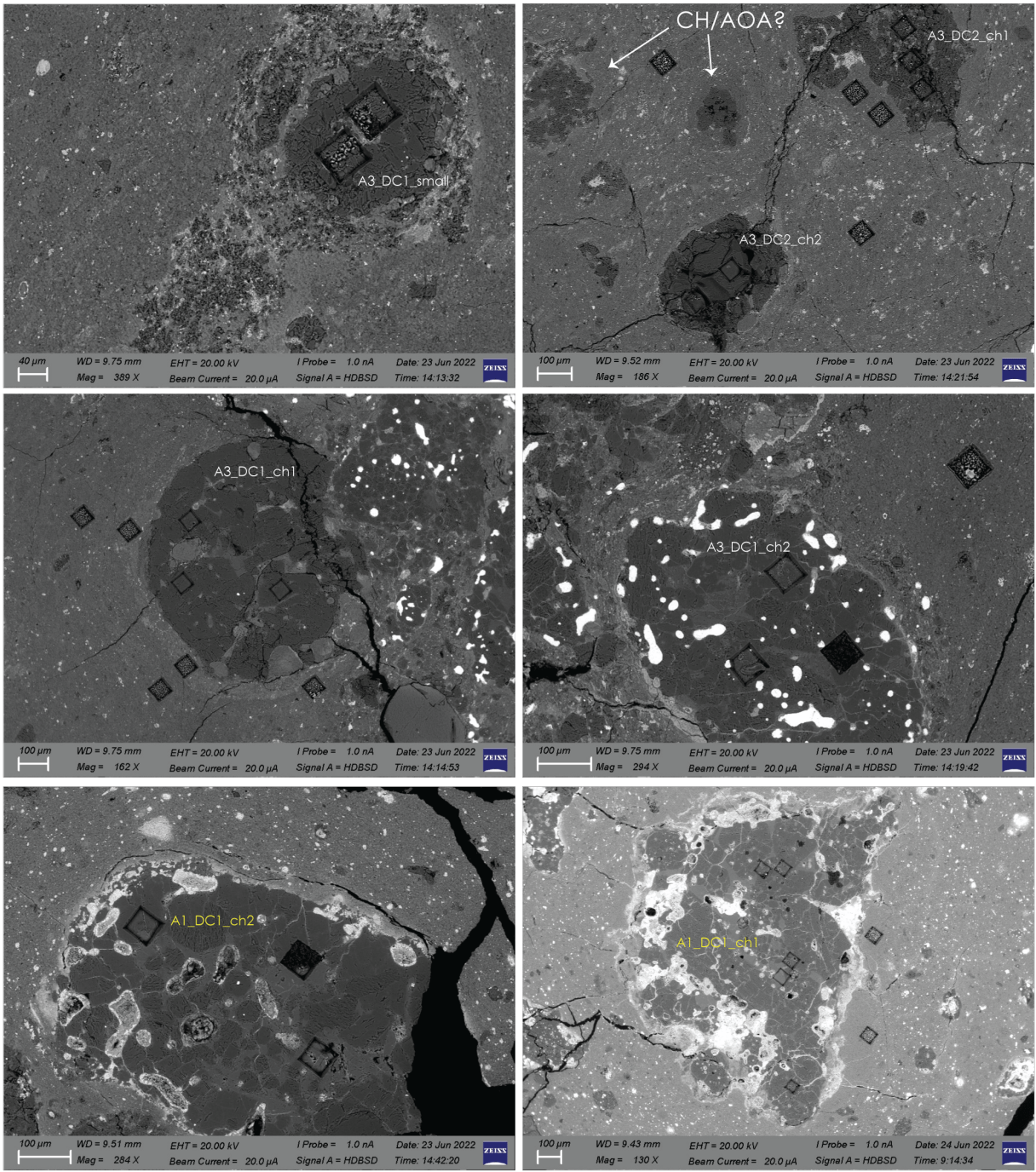

**Fig. S6:** Laser ablation pits from chondrule analyses in dark clasts (see Data S1). CH: chondrules, AOA: amoeboid olive aggregate.

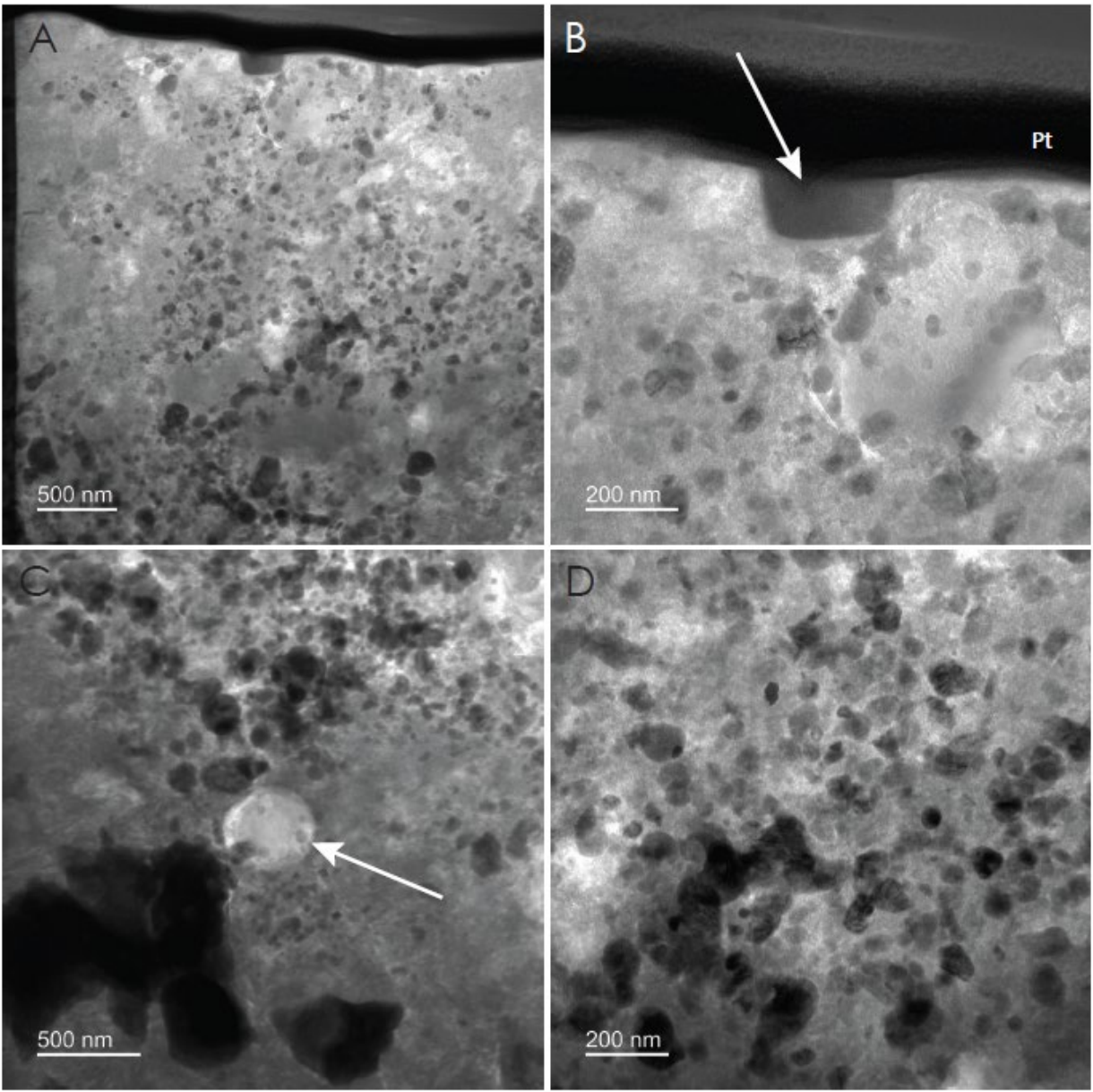

**Fig. S7:** Transmission electron microscopy (TEM) section (C2_3) in bright field (BF) mode from dark clast C2DC5, where we fibbed out the presolar O-anomalous grain and the surrounding $^{15}N$ and D-rich organic matter (see also main text). **A)** Overview of TEM section showing black grains (sulfides), white (organics) and grey matter (silicates). **B)** Presolar grain (arrow) surrounded by amorphous silicates, sulfides, and organic matter. **C)** top of figure: sulfide-organic complexes similar to glassy embedded metal and sulfide (GEMS)-like materials, middle: globular organic grain (arrow); left corner: phyllosilicate structure and larger sulfide grains points to a more altered part of the TEM section. **D)** Zoom-in on sulfide-organic-complex (SOC) region.

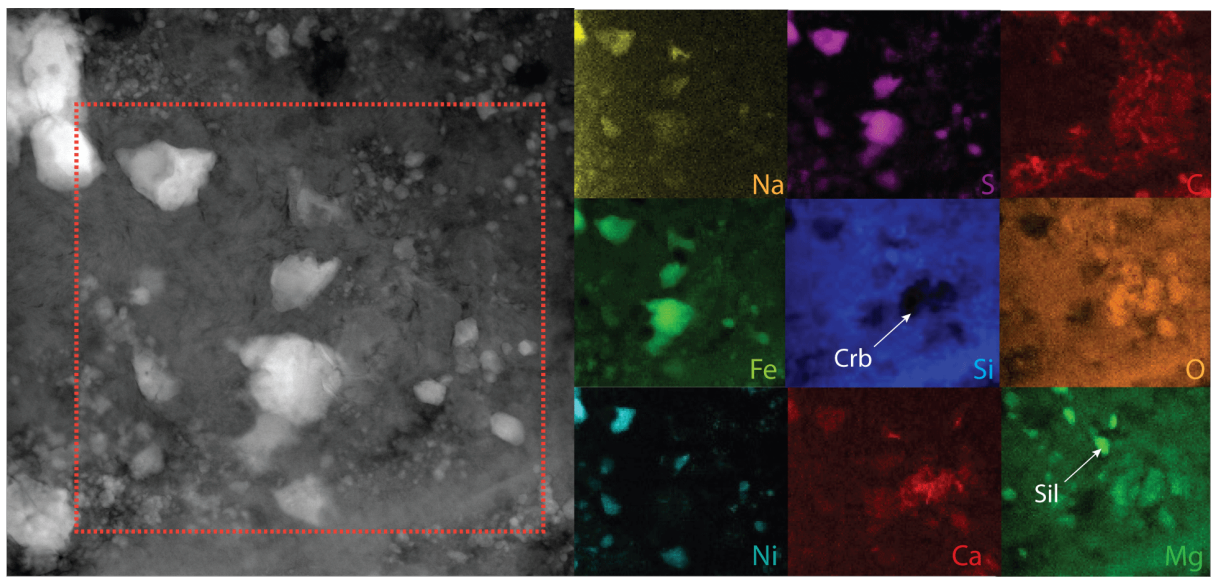

**Fig S8:** High-angle annular dark-field STEM imaging (HAADF) of region of interest (ROI) in section C2_3 with red box indicating area of energy dispersive spectroscopy (EDS) mapping (see right colored K-edge elemental maps). Sulfide grains (white in HAADF) are dominantly FeS, but some individual grains are Fe-poor, Ni-rich and some grains also appear to incorporate significant Na, although the latter might reflect overlap with the Zn-K peak. We have also identified Ca- and Mg-rich carbonates (Crb) in the organosulfide-rich areas, as well as Mg-rich, Fe-poor silicates grains (Sil).

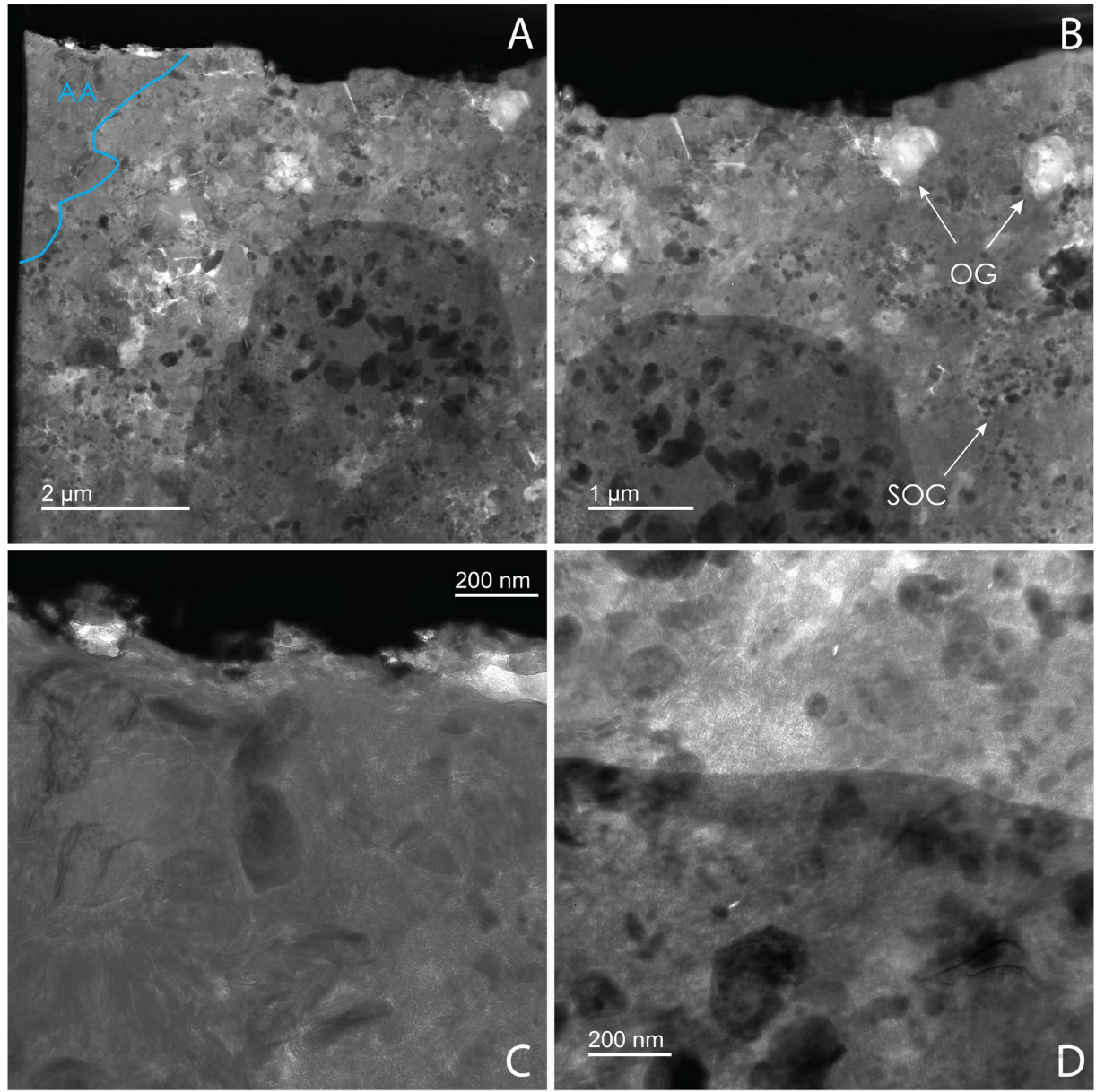

**Fig. S9: A) overview of section C2_2 from clast C2DC5 in bright field (BF) mode**. Left upper corner is more aqueously altered (AA zone), with larger phyllosilicates and altered sheeted sulfides (see zoom-in from panel **C**). The section contains a rounded 'darker' zone in the middle that is mineralogically similar to the rest of the section and at the edges contains grains that continue past the boundaries of this zone (**D**). The edges of this zone are sharper at the top and fade out at the bottom. In and outside the dark zone the silicates are amorphous and there is no indication that the dark zone is more or less aqueously altered. **B)** Outside of the dark zone are abundant sulfide-organic complexes (SOC) and globular organic matter (OG).

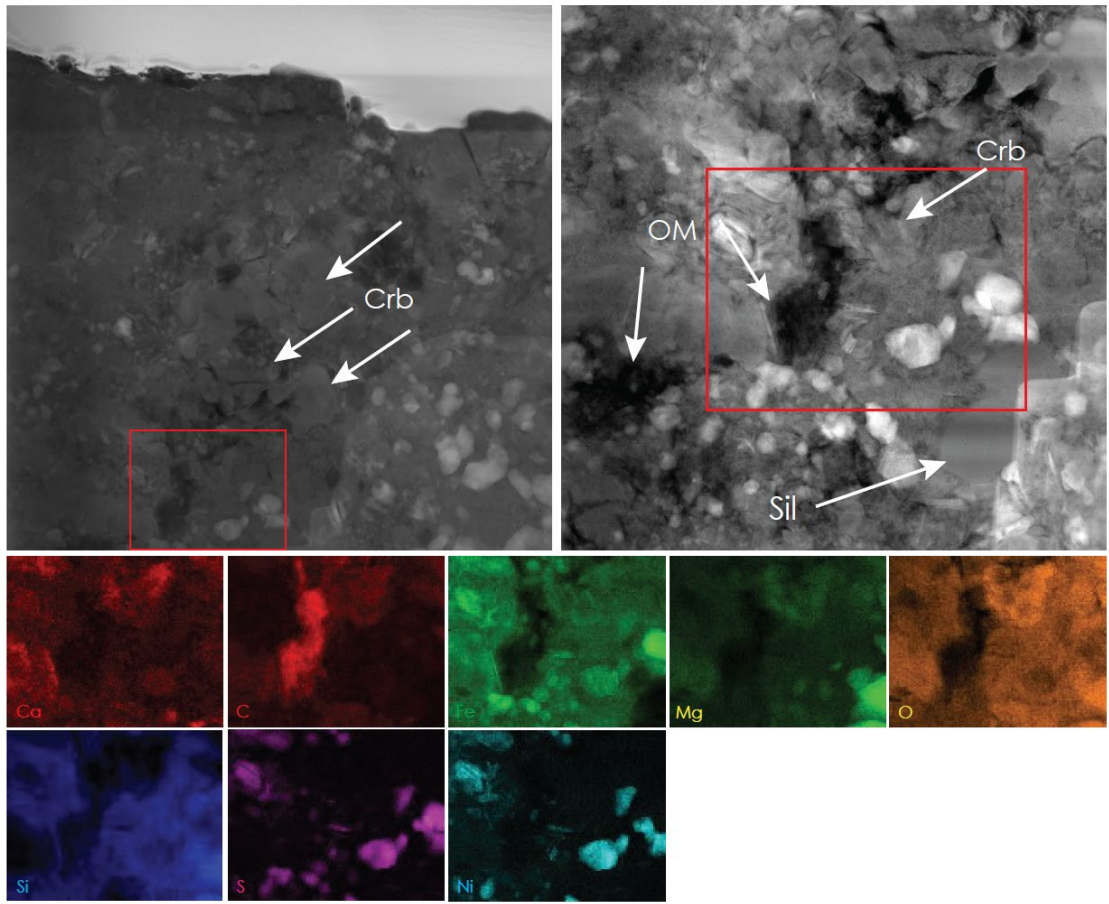

**Fig S10:** Zoom-in of section C2_2 on a region rich in sulfide-organic complexes (SOC) and neighboring the dark zone in scanning transmission electron microscopy (STEM)- high-angle annular dark-field (HAADF) mode. Red box indicates the location of energy dispersive spectroscopy (EDS) mapping. Within this region there are abundant Ca- and Mg-rich carbonates, similar to those observed in section C2_3. Sil = Mg-rich silicate grain. Crb = carbonate.

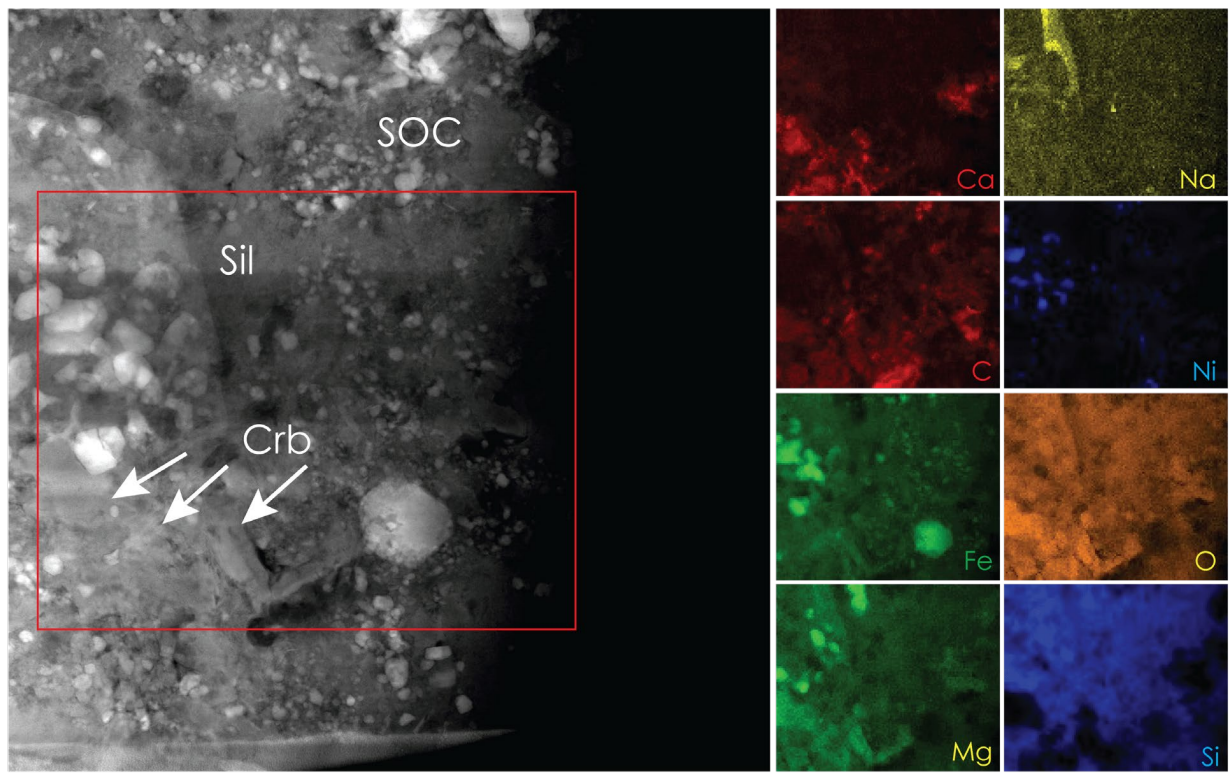

**Fig S11:** Close-in on the boundary between the dark zone and the rest of the section. Red box indicates the area with energy dispersive spectroscopy (EDS) mapping (see colored panels). The sharp boundary is marked by an enrichment in Na and the dark zone area appears to be enriched in Mg relative to the surrounding area. The dark zone also has more Ni-rich sulfides than the other regions. Sil = Mg-rich silicate grain. Crb = carbonate.

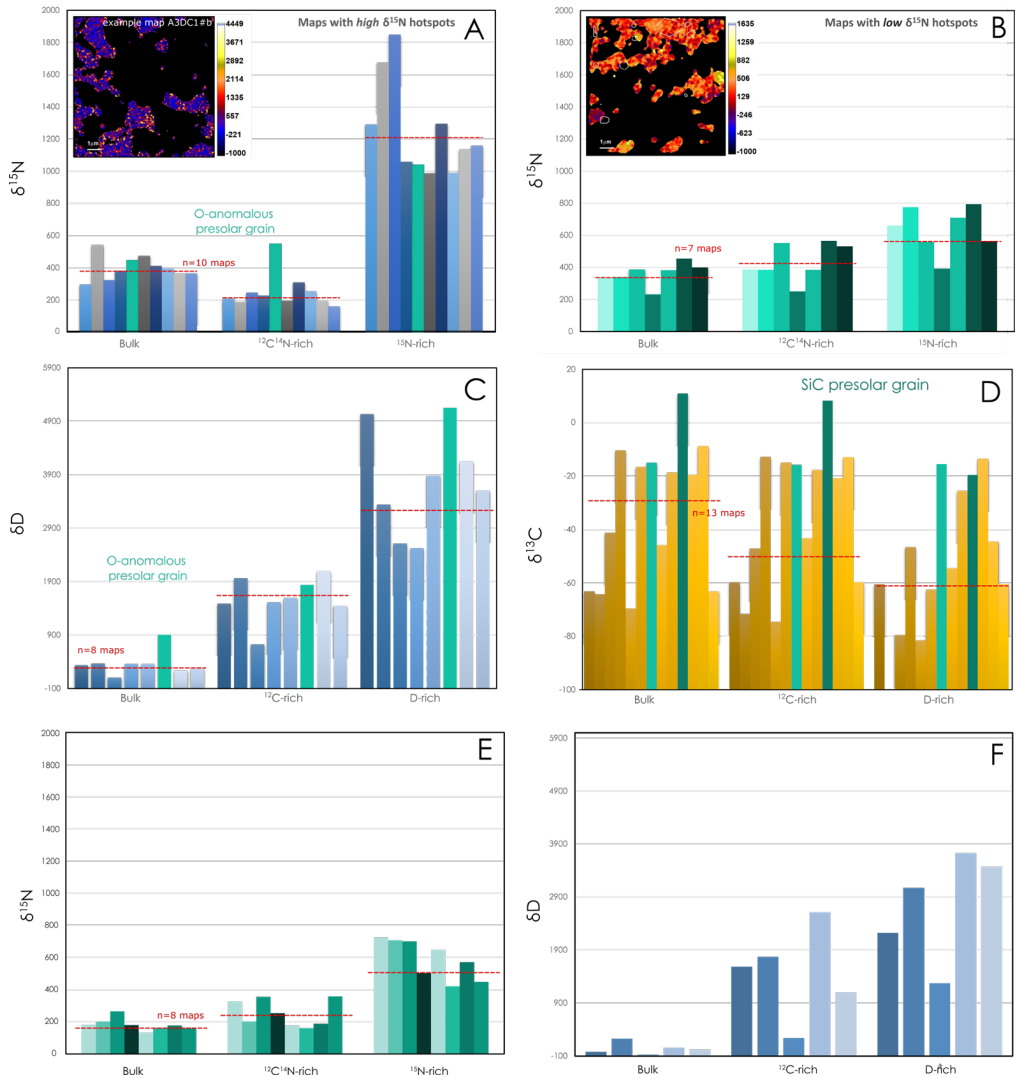

**Fig. S12: Nitrogen, hydrogen and carbon isotope data from nano secondary ion mass spectrometry analyses on selected matrix areas from dark clasts**. See Data S2 for map locations, data reduction methods and map isotope images. **A)** Nitrogen isotope data from 10 individual isotope maps with relatively high $^{15}$N-rich hotpots ($\delta^{15}$N >1000 ‰), including bulk maps, $^{12}$C$^{14}$N-rich areas (i.e., organic matter, using a 20% lower counts cut-off mask) and $^{15}$N-rich areas (using a mask of the most enriched 20%). Red dashed lines depict averages that are used in Fig. 3 of the main text. The accompanying isotope map shows the distribution of $^{15}$N, with the most $^{15}$N-rich areas plotting at the edges of the carbon-rich areas. **B)** Same as A but for 7 maps with low $^{15}$N-rich hotpots ($\delta^{15}$N <800 ‰). The accompanying isotope map shows the distribution of $^{15}$N, with the most $^{15}$N-rich areas plotting more homogeneously within the carbon-rich areas. We observe two trends: A) decrease in $^{15}$N with increasing carbon content and B) increase in $^{15}$N with increasing carbon content. See supplementary text for interpretation. **C)** Hydrogen isotope data from 8 individual maps, including bulk maps, $^{12}$C-rich areas (i.e., organic matter, using a 20% lower counts cut-off mask) and D-rich areas (using a mask of the most enriched 2%). **D)** Carbon isotope data analyzed simultaneously with hydrogen isotopes. **E)** Nitrogen isotope compositions of 8 individual maps from dark clast C3DC4. **F)** Hydrogen isotope compositions of dark clast C3DC4.

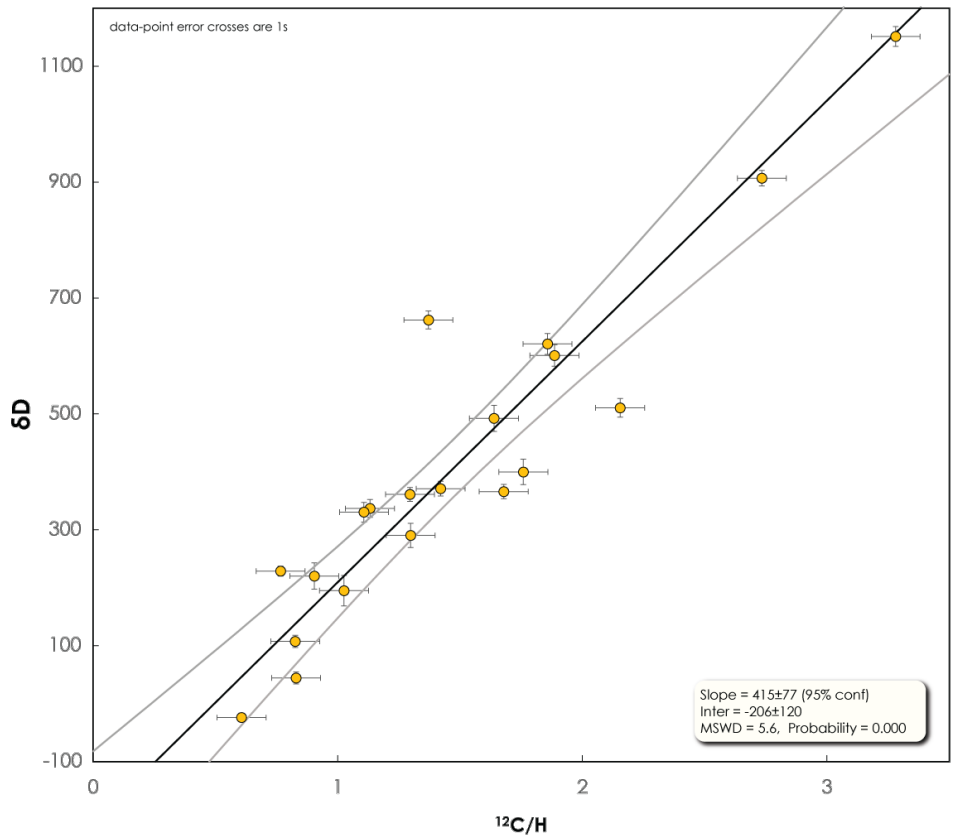

**Fig. S13:** The raw $^{12}$C/H ratios from nano secondary ion mass spectrometry analyses of dark clast bulk and $^{12}$C-rich compositions versus the corrected δD values. The correlation is obtained using a model 2 Yorkfit in Isoplot4.15. The δD value at the intercept of this correlation is suggested to represent the initial water ice composition of the clasts (*33*). This value of −206±120 ‰ (2SD) corresponds roughly to the average D/H ratio of initial water found for carbonaceous chondrites (CC) (δD$_{CM}$ = −350±40 ‰, δD$_{CI}$ ≈ +100 ‰) (*33*), including Isheyevo lithic clasts (δD = −350 ‰, obtained from a correlation between δD and Si/H) (*31*). The very similar D/H ratios of CC initial water suggests efficient equilibration of interstellar medium derived icy grains with H$_2$ in the protoplanetary disk (*101*). Note that the instrumental mass fractionation (IMF) corrected C/H ratios by mass (2.5±1.2) are given in Data S2 and are in agreement with literature values for carbonaceous chondrites (*102*). The uncertainty of the IMF corrected C/H values has been suggested to be smaller than a factor 2 (*87*), and increasing the error would steepen the slope of the regression line.

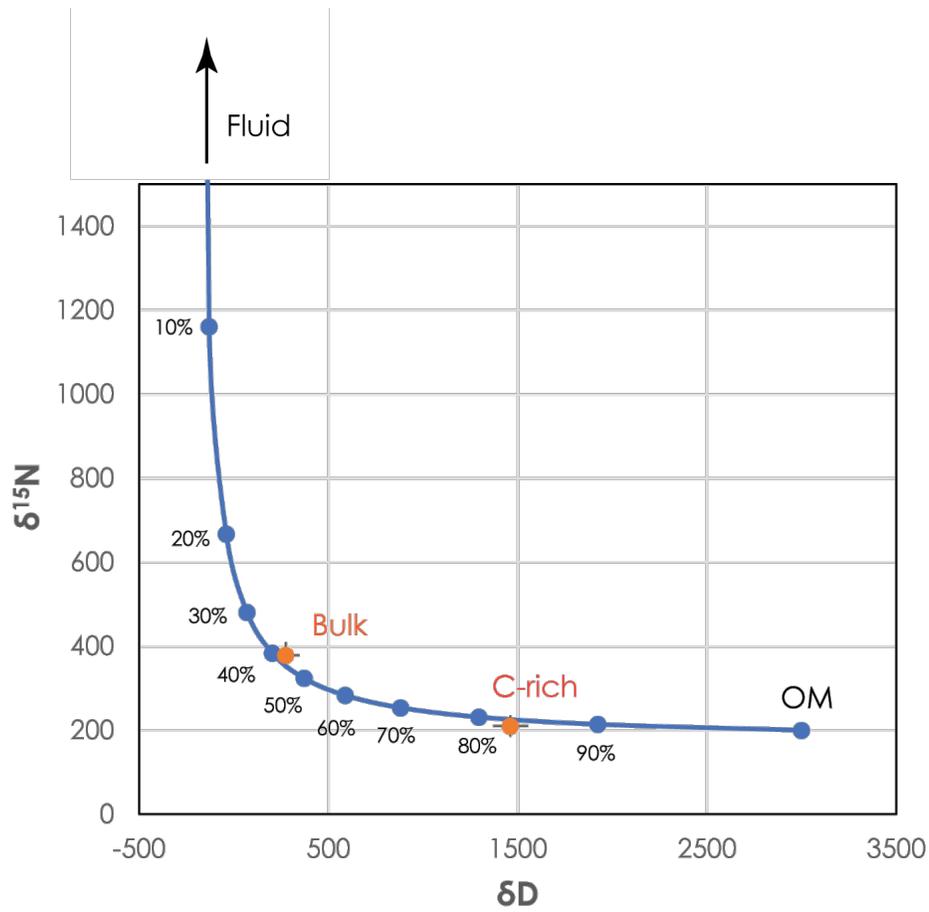

**Fig. S14:** Optimal mixing line calculated for the bulk and C-rich areas of the dark clasts in $\delta D$ and $\delta^{15}N$ space. See also Table S2 for model parameters. From this diagram we calculate 40% contribution of organic matter (OM) and 60% contribution from the initial ice (fluid). This ratio is compared against the organic: silicate ratio we obtain from transmission electron microscopy petrography (Table S3).

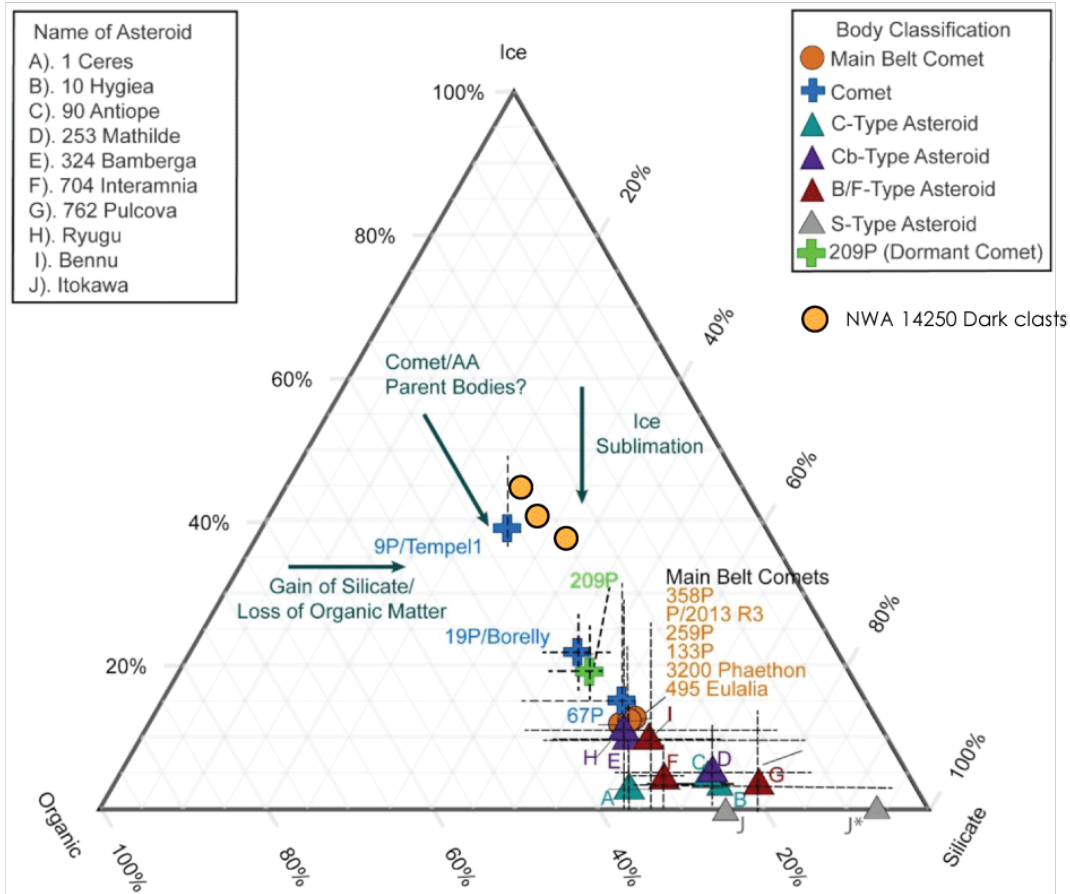

**Fig. S15:** Ternary plot with ice, organics and silicates as endmember compositions, modified after Havishk et al. (*41*). This plot shows the very pristine nature of the dark clasts, which is similar to comets, but clearly different from asteroids.

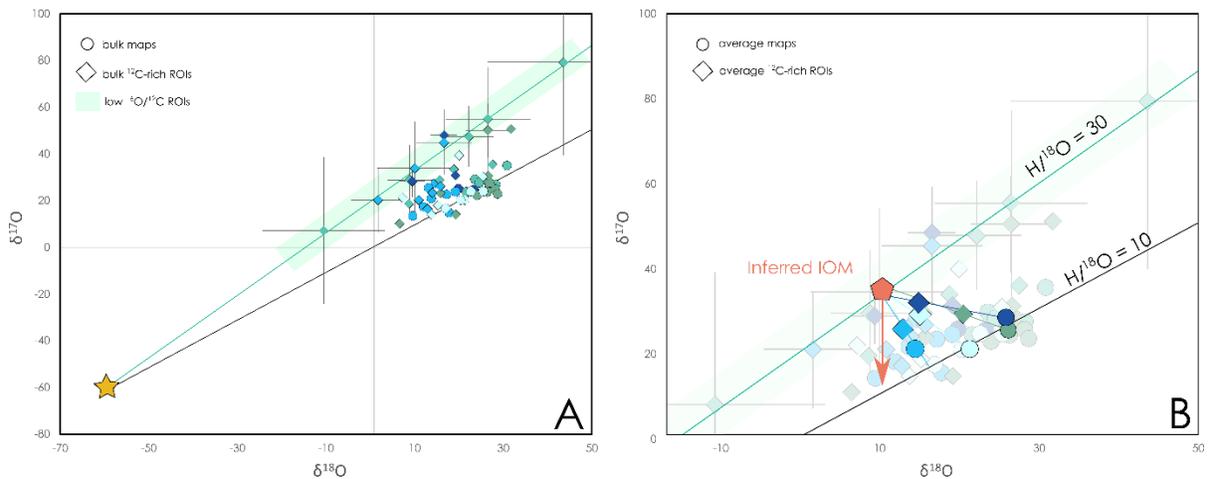

**Fig. S16: Three oxygen isotope plot** including **A)** the Young and Russell slope 1 line (solid black) with and intercept at the solar composition (star) (*42*). Circles are bulk isotope maps from the dark clasts, with 1σ propagated uncertainties (from counting statistics and San Carlos olivine standard) for δ$^{17}$O and δ$^{18}$O typically better that 3 ‰ and 1.5 ‰, respectively. The diamonds show bulk $^{12}$C-rich areas from individual maps, which typically have large uncertainties due to a decrease in oxygen counts relative to the silicon-rich areas. Regions of interest that are particularly $^{12}$C-rich, with very low $^{16}$O/$^{18}$O ratios (highlighted green in panel A), form a correlation that also intercepts with the sun, but do not fall on a slope 1 line that is suggested to reflect self-shielding processes. In panel **B**, we show that averaging the individual bulk isotope maps for each dark clast (n = 4) and doing the same for the $^{12}$C-rich areas, we can draw mixing lines between the bulk and the $^{12}$C-rich areas. All mixing lines for these clasts point to the same organic endmember, which we suggest represents the bulk insoluble organic matter (IOM) of the dark clasts. However, one issue with the analyses of areas with high H/O ratios is the effect of $^{16}$OH- tailing on the $^{17}$O peak, which is only resolved at very high mass resolving power (*103*). Hence, it is most likely that a correction needs to be made to the IOM composition, which would place it closer to or on the slope 1 line, similar to CI chondrite IOM (*103*).

| Type II chondrule ferrous olivine | Cr (ppm) | $1\sigma$ |
|---|---|---|
| C$_5$_OB1A_2 | 3250 | 120 |
| C$_5$_OB1A_4 | 2539 | 79 |
| C$_5$_OB1A_5 | 3650 | 190 |
| C$_5$_OB1A_8 | 3058 | 96 |
| C$_5$_OB1A_9 | 3350 | 140 |
| | | |
| C$_5$_OB2_2 | 2082 | 63 |
| C$_5$_OB2_4 | 1776 | 56 |
| C$_5$_OB2_5 | 2416 | 70 |
| | | |
| C$_3$_OB1_1 | 3260 | 110 |
| C$_3$_OB1_5 | 2735 | 88 |
| C$_3$_OB1_6 | 3490 | 100 |
| | | |
| C$_3$_OB2_1 | 2620 | 100 |
| C$_3$_OB2_2 | 2552 | 94 |
| C$_3$_OB2_3 | 2335 | 69 |
| C$_3$_OB2_4 | 2738 | 75 |
| | | |
| **average** | **2790** | **540** |
| Cr$_2$O$_3$ (ppm) | **4078** | **789** |

**Table S1:** Chromium analyses of ferrous olivine grains from type II chondrules in CR chondrite NWA 14250.

| | N/C (at) | H/C (at) | N/H (at) | H/N (at) IOM |
|---|---|---|---|---|
| CR chondrites* | 3.39 | 78.3 | 0.043 | 23.1 |
| | 3.65 | 80.3 | 0.045 | 22.0 |
| | 3.83 | 75.7 | 0.051 | 19.8 |
| | 4.39 | 68.9 | 0.064 | 15.7 |
| | 3.21 | 80.6 | 0.040 | 25.1 |
| | 3.16 | 78.6 | 0.040 | 24.9 |
| | 3.19 | 71.9 | 0.044 | 22.5 |
| **CR average** | **3.55** | **76.33** | **0.047** | **21.9** |

| | H:N (at) fluid | X1 | X2 | $\delta^{15}N$ mix | $\delta D$ mix |
|---|---|---|---|---|---|
| $NH_3$ | 3 | | | *200* | *3000* |
| HCN | 1 | 0.1 | 0.9 | 214 | 1924 |
| HCN : $H_2O$ (1:1) | 3 | 0.2 | 0.8 | 231 | 1295 |
| $NH_3$ : $H_2O$ (1:1) | 5 | 0.3 | 0.7 | 254 | 881 |
| $NH_3$ : $H_2O$ (1:10) | 23 | 0.4 | 0.6 | 283 | 589 |
| $NH_3$ : $H_2O$ (1:500) | 1000 | 0.5 | 0.5 | 324 | 372 |
| | | **0.6** | **0.4** | **383** | **204** |
| | | 0.7 | 0.3 | 481 | 70 |
| | | 0.8 | 0.2 | 667 | -39 |
| | | 0.9 | 0.1 | 1161 | -130 |
| | | | | *6500* | *-206* |

| | | Fluid [xs]1 | OM [xs]2 | Fluid e 1 | OM e 2 |
|---|---|---|---|---|---|
| $\delta^{15}N$ | | 0.02 | 1 | 6500 | 200 |
| $\delta D$ | | 100 | 22 | -206 | 3000 |

**Table S2:** Average composition and H:N ratio (21.9) of insoluble organic matter (IOM) from CR chondrites analyzed by Alexander et al. (*56*). Parameters that constrain the mixing line calculations, including the H:N ratios [xs] and $\delta D$ and $\delta^{15}N$ compositions (e1 and e2) of fluid and organic matter. X1 and X2 are respectively fractions of endmember 1 and 2. Red font refers to the endmember concentrations for the nitrogen and hydrogen isotope values observed for the bulk composition.

|  | ice | organic | silicate | sum |
|---|---|---|---|---|
| *ratio* | 1.5 | 1 | 1 | 3.5 |
| *percentage* | 43 | 29 | 29 | |
| *ratio* | 1.5 | 1 | 1.2 | 3.7 |
| *percentage* | 41 | 27 | 32 | |
| *ratio* | 1.5 | 1 | 1.4 | 3.9 |
| *percentage* | 38 | 26 | 36 | |

**Table S3:** The ice: organic: silicate ratios of dark clasts based on ice: organic ratios from Fig. S14 and organic: silicate ratios from transmission electron microscopy petrography (see main text). Based on the uncertainties of the latter, a range is given towards the most conservative ratio estimates.

| | Comment | $\delta^{56}$Fe | 2se | $\mu^{54}$Fe | 2se | $^{27}$Al/$^{24}$Mg | $\delta^{25}$Mg | 2SE | $\mu^{26}$Mg* | 2se |
|---|---|---|---|---|---|---|---|---|---|---|
| *Dark clasts* | | | | | | | | | | |
| A3_DC1 | confirmed ODD | -0.045 | 0.004 | 34 | 7 | 0.098 | 0.030 | 0.009 | -15.7 | 1.8 |
| C2C3 | confirmed ODD | -0.038 | 0.007 | 32 | 15 | 0.111 | -0.034 | 0.020 | -15.8 | 4.5 |
| **average ODDs** | | **-0.042** | **0.010** | **33** | **3** | **0.105** | **-0.002** | **0.091** | **-15.8** | **0.1** |
| | | | | | | | | | | |
| A1 | not confirmed | -0.042 | 0.010 | 26 | 4 | 0.095 | -0.030 | 0.011 | -9.6 | 2.2 |
| A2 | heated clast | -0.023 | 0.009 | 26 | 4 | 0.381 | -0.009 | 0.013 | -6 | 3.9 |
| | | | | | | | | | | |
| *Chondrites* | | | | | | | | | | |
| CI* | | 0.060 | 0.030 | -2 | 3 | 0.096 | -0.015 | 0.038 | 4.5 | 0.4 |
| CR* | | 0.050 | 0.070 | 29 | 4 | 0.145 | -0.042 | 0.102 | -4.7 | 3.9 |
| | | | | | | | | | | |
| BIR-1 | ref std | | | | | | 0.011 | 0.009 | -3.1 | 2.6 |

| | Comment | $^{55}$Mn/$^{53}$Cr | $\delta^{53}$Cr | 2se | $\mu^{53}$Cr | 2se | $\mu^{54}$Cr | 2se |
|---|---|---|---|---|---|---|---|---|
| *Dark clasts* | | | | | | | | |
| A3_DC1 | confirmed ODD | 0.84 | -0.29 | 0.01 | -12 | 5 | 125 | 9 |
| C2C3 | confirmed ODD | 0.93 | | | | | | |
| **average ODDs** | | **0.89** | -0.29 | 0.01 | -12 | 5 | **125** | **9** |
| | | | | | | | | |
| A1 | not confirmed | 0.62 | -0.22 | 0.01 | 24 | 4 | 131 | 5 |
| A2 | heated clast | 0.72 | -0.20 | 0.02 | -9 | 8 | 180 | 13 |
| | | | | | | | | |
| *Chondrites* | | | | | | | | |
| CI* | | 0.85 | -0.15 | 0.01 | 23 | 5 | 162 | 16 |
| CR* | | 0.79 | -0.14 | 0.03 | 19 | 7 | 122 | 22 |

**Table S4:** Fe, Mg, and Cr isotope data for several dark clasts. Mass-dependent isotope data is depicted in the delta notation (as parts per thousand relative to a standard, see Methods) and mass-bias corrected data in the mu notation (as parts per million). All isotope data are obtained using a Neptune Plus multi-collector inductively coupled mass spectrometer (MC-ICPMS) at StarPlan (University of Copenhagen). Elemental ratios were obtained using a ThermoFisher iCAP ICPMS at StarPlan. Of the four measured clasts, two are confirmed outer disk dark clasts (ODDs), based on in situ nano secondary ion mass spectrometry analyses. Note that C2C3 is a combined aliquot of two smaller dark clasts that could not be analyzed individually and was too small for chromium isotope analyses. Clast A2 has been briefly heated and has an elevated $^{27}$Al/$^{24}$Mg ratio relative to the solar value of 0.096. For comparison, we also show isotope data for CI and CR chondrites (*7, 10, 48, 50*).

| | $\partial^{30}Si$ (%) | 2SE | $\partial^{29}Si$ (%) | 2SE | $\mu^{30}Si$ | 2SE | $\mu^{29}Si$ | 2SE |
|---|---|---|---|---|---|---|---|---|
| **A3DC1** | -0.35 | 0.02 | -0.18 | 0.01 | -2.8 | 7.2 | 1.4 | 3.7 |
| | -0.37 | 0.04 | -0.18 | 0.02 | 10.2 | 5.6 | -5.2 | 5.8 |
| | -0.37 | 0.04 | -0.19 | 0.02 | 2.0 | 9.0 | -1.0 | 4.6 |
| | -0.34 | 0.01 | -0.18 | 0.00 | 6.8 | 4.0 | -3.4 | 2.0 |
| average | -0.36 | 0.01 | -0.18 | 0.00 | 4.1 | 5.7 | -2.1 | 2.9 |
| *wt. mean* | | | | | **5.7** | **2.8** | | |
| | | | | | | | | |
| **A3DC2** | -0.35 | 0.01 | -0.18 | 0.01 | 3.0 | 3.2 | -1.4 | 1.6 |
| | -0.38 | 0.04 | -0.19 | 0.02 | 8.3 | 7.4 | -4.2 | 3.8 |
| | -0.36 | 0.04 | -0.19 | 0.02 | 6.9 | 9.0 | -3.5 | 4.6 |
| | -0.38 | 0.02 | -0.19 | 0.01 | 2.6 | 5.5 | -1.3 | 2.8 |
| average | -0.37 | 0.01 | -0.19 | 0.00 | 5.2 | 2.8 | -2.6 | 1.5 |
| *wt. mean* | | | | | **3.8** | **2.4** | | |
| | | | | | | | | |
| *All cometary clasts* | | | | | | | | |
| *wt. mean* | | | | | **4.7** | **1.8** | | |
| | | | | | | | | |
| Á1DC1 | -0.47 | 0.001 | -0.02 | 0.004 | 15.8 | 5 | -8 | 2.5 |
| | | | | | | | | |
| *Chondrites* | | | | | | | | |
| CI* | -0.48 | 0.01 | | | 32.0 | 2.0 | | |
| CR* | -0.48 | 0.11 | | | 10.0 | 4.0 | | |
| | | | | | | | | |
| *Standard* | | | | | | | | |
| BHVO-2 | -0.28 | 0.04 | -0.13 | 0.01 | -0.5 | 7.9 | -0.2 | 4.0 |

**Table S5:** Si isotope data for two confirmed outer disk dark clasts (ODDs: A3DC1 and A3DC2) and CR-like clast A1DC1. A3DC1 is also analyzed for iron, magnesium and chromium isotopes (see Table S4). Mass-dependent isotope data is depicted in the delta notation (as parts per thousand relative to a standard, see Methods) and mass-bias corrected data in the mu notation (as parts per million). All isotope data are obtained using a Neptune Plus multi-collector inductively coupled mass spectrometer at StarPlan (University of Copenhagen). For comparison, we also show isotope data for CI and CR chondrites (*49*). A BHVO-2 reference standard was processed and analyzed along with the dark clasts.

**Data S1:** Laser ablation inductively coupled mass spectrometer data of individual spots on dark clasts in NWA 14250.

**Data S2:** Nano secondary ion mass spectrometry (nanoSIMS) data including H, N, C and O isotope compositions of individual isotope maps in dark clasts from NWA 14250. We also provide data reduction methods and nanoSIMS map locations using back scattered electron images of the dark clasts.